\documentclass[11pt ]{revtex4}
\usepackage{hyperref}
\usepackage{amsmath,amsfonts}
\usepackage[dvips]{graphics}
\usepackage[dvips]{graphicx}
\usepackage{subfigure}
\usepackage{epsfig}
\usepackage{epstopdf}
\usepackage{amsmath}
\usepackage{amsfonts}
\RequirePackage{color}

\begin{document}
\title{\Large Perturbed thermodynamics of charged black hole solution in Rastall theory}
  \author{Behnam Pourhassan${}^{a,b}$}
  \email{b.pourhassan@du.ac.ir}
  \author{Sudhaker Upadhyay${}^{c,d,e,a}$}
  \email{sudhakerupadhyay@gmail.com; sudhaker@associates.iucaa.in}
   \affiliation{${}^{a}$School of Physics, Damghan University, P. O. Box 3671641167, Damghan, Iran.}
   \affiliation{${}^{b}$Canadian Quantum Research Center 204-3002 32 Avenue Vernon, British Columbia V1T 2L7 Canada.}
 \affiliation{${}^{c}$ Department of Physics, K. L. S. College,  Nawada-805110,  India.}
 \affiliation{${}^{d}$Department of Physics, Magadh University, Bodh Gaya, Bihar 824234, India}
  \affiliation{${}^{e}$ Visiting Associate, Inter-University Centre for Astronomy and Astrophysics (IUCAA),  Pune, Maharashtra-411007, India. }

\begin{abstract}
We study the thermodynamics of a charged black hole surrounded by a perfect fluid in Rastall theory and investigate three different cases of quintessence, dust and radiation fields. By considering  thermal fluctuations, we study the corrected thermodynamics of the system. We investigate the effects of thermal fluctuations on the black hole stability, phase transition and critical points. We show that the thermal fluctuations increase instability of the black hole and may yield  to the second order phase transition. We also compare our result  with uncharged cases to find
the effects of the black hole charge on the thermodynamics quantities. We find that large black holes behave like a Van der Waals fluid, while for the small black hole where thermal fluctuations become important, there is no Van der Waals behaviour. Finally, we briefly discuss about geometric thermodynamics to obtain corrected scalar curvature.
\end{abstract}
\maketitle
\section{Introduction}
In order to explain an accelerating phase of the universe, a   modification to Einstein general relativity (GR) is proposed  by
introducing the coupling between matter and gravitational fields in a non-minimal
way, which violets energy-momentum conservation in the curved space-time \cite{ras1, ras2}.
In this modified GR, the covariant divergence of energy-momentum tensor should depend on the space-time curvature through a coupling parameter so that  energy-momentum conservation can be recovered in the weak gravitational field limit or Minkowski flat space-time. In this consideration,  the Einstein's   equation gets modified to
\begin{equation}
G_{\mu\nu}+\kappa\lambda g_{\mu\nu}R=\kappa T_{\mu\nu},
\end{equation}
 where $\lambda$ is called the Rastall parameter. This field equation has found
many exact solutions for both, astrophysical and cosmological frameworks \cite{2,3,4,5,6,7,8}.
In recent years, Rastall theory of modified GR has found great attention to many researches
\cite{9,10,sud0}. For example, Rastall field equations are investigated by using a Lagrangian formalism \cite{Ras1, Ras2}.

 On the other hand, black holes are very important objects for the perspectives of astrophysics, cosmology and particle physics. The rotating black hole in Rastall theory has  already been studied by the Ref. \cite{kumar}. One of the best way to obtain information about black holes is thermodynamics study. In this regard, we found that thermodynamics of black holes in Rastall gravity is studied by the Ref. \cite{morad222} and obtained that there is a lower bound for the horizon radius which satisfied by black holes stability condition.

One of the important issue in the black hole thermodynamics is consideration of the thermal fluctuations \cite{das, rahim}. Thermal fluctuations are due to the statistical perturbations which may be interpreted as quantum effects. Because these are important when the black hole size decreases due to the Hawking radiation \cite{Hawking1, Hawking2}.  It means that when the black hole will be sufficiently small, then its temperature will be so large and thus one can not ignore the significance of thermal fluctuations \cite{NPB1}. Investigation of thermal correction effects in a strong gravitational system is so important and interesting. Hence, study of thermal fluctuations in a given black hole help us to understand  the theory of gravity. Recently, this subject takes more attention so there are several works where thermal fluctuations of a given black hole have been studied \cite{sud1,sud2,sud3,sud4,sud5,sud6,sud7,sud8,sud9,hay}. Now, we would like follow Ref. \cite{sud0} to study thermal fluctuations on thermodynamics of the black hole surrounded by perfect fluids in Rastall theory of gravity.

We present this paper in following manner. In section II, we discuss a  black hole solution 
surrounded by a perfect fluid in Rastall theory. Also, we shed light on the thermodynamics of this black hole solution in three special cases of
perfect fluid, namely, quintessence, dust and radiation field. In section III,
we give details of corrected entropy  due to small statistical disturbance around equilibrium.
Section IV is devoted to study the effects of thermal fluctuation on the various thermodynamical equations of state. The critical points and stability of such black hole are
emphasized in section V. In section VI, we briefly discuss about geometrothermodynamics to obtain thermal fluctuation effects on the black hole stability. Finally, we summarize results of this paper in the last section.

\section{Black hole solution}
The metric of a charged black hole with mass $M$ and charge $Q$ surrounded by 
perfect fluid in Rastall theory with Rastall gravitation coupling constant $k
$ and $\lambda$ is given by \cite{HD} 
\begin{equation}\label{1}
ds^{2}=-\left(1-\frac{2M}{r}+\frac{Q^{2}}{r^{2}}-\frac{N_{s}}
{r^{\Lambda_{s}}}\right)dt^{2}
+\frac{dr^{2}}{\left(1-\frac{2M}{r}+\frac{Q^{2}}{r^{2}}-\frac{N_{s}}
{r^{\Lambda_{s}}}\right)}
+r^{2}d\Omega^{2},
\end{equation}
where $N_s$ denotes the surrounding field structure parameter with the equation of state parameter $\omega_{s}$  and
\begin{equation}\label{2}
\Lambda_{s}=\frac{1+3\omega_{s}-6k\lambda(1+\omega_{s})}{1-3k\lambda(1+\omega_{s})}.
\end{equation}
The special case of $\lambda=0$ and $k=8\pi G$ reduces this to the Reissner-Nordstr\"{o}m black hole surrounded by a
surrounding field \cite{Kiselev}.

The black hole event horizon $r_{+}$ is given by the largest root of the following relation:
\begin{equation}\label{3}
1-\frac{2M}{r}+\frac{Q^{2}}{r^{2}}-\frac{N_{s}}{r^{\Lambda_{s}}}=0.
\end{equation}
In that case the black hole entropy is given by 
\begin{equation}\label{4}
S=\pi r_{+}^{2}.
\end{equation}
Also, the black hole temperature is obtained by the following relation:
\begin{eqnarray}\label{5}
T&=&\frac{1}{4\pi}\frac{d}{dr}\left(1-\frac{2M}{r}+\frac{Q^{2}}{r^{2}}-\frac{N_{s}}{r^{\Lambda_{s}}}\right)_{r=r_{+}} \nonumber\\
&=&\frac{1}{4\pi}\left(\frac{2M}{r_+^2}-2\frac{Q^2}{r_+^3}+\Lambda_s\frac{N_s}{r_+^{\Lambda_s+1}}\right).
\end{eqnarray}
By using the equation (\ref{3}), the black hole mass is calculated as
\begin{equation}\label{6}
M=\frac{r_{+}}{2}+\frac{Q^{2}}{2r_{+}}-\frac{N_{s}}{2r_{+}^{\Lambda_{s}-1}}.
\end{equation}
Also chemical potential (electrostatic potential) conjugated with the black hole charge is given by,
\begin{equation}\label{chemical}
\Phi=\frac{Q}{r_{+}}.
\end{equation}
The first law of black hole thermodynamics is given by 
\begin{equation}\label{1l}
dM=TdS+\Phi dQ+\Theta dN_{s},
\end{equation}
where the generalized force corresponding to the surrounding field structure parameter $N_{s}$ is
\begin{equation}\label{force}
\Theta=-\frac{1}{r_{+}^{\Lambda_{s}-1}}.
\end{equation}
 We can find that the first law of thermodynamics satisfies when  the following condition holds:
\begin{equation}\label{con}
Q=\left[\int{(r_{+}^{2-\Lambda_{s}}\frac{dN_{s}}{dr_{+}})dr}+c\right]^{\frac{1}{2}},
\end{equation}
where $c$ is an integration constant. It is clear that for the constant $Q$ and $N_{s}$ (which is our consideration) the first law of thermodynamics satisfies.

\subsection{Quintessence field}
The black hole surrounded by the quintessence field  is given by setting $\omega_{s}\equiv\omega_{q}=-\frac{2}{3}$ or $\omega_{q}=-\frac{1}{3}$ \cite{kumar}. Both of them yield  to the similar results in our calculations. Hence, we consider the case of $\omega_{q}=-\frac{2}{3}$ only.
In the case of $k\lambda=\frac{1}{4}$ \cite{sud0}, we have $\Lambda_{s}\equiv\Lambda_{q}=-2$. In this case, the equation (\ref{3}) reduces to the following form:
\begin{equation}\label{7}
1-\frac{2M}{r}+\frac{Q^{2}}{r^{2}}-N_{q}r^{2}=0.
\end{equation}
\begin{figure}[h!]
 \begin{center}$
 \begin{array}{cccc}
\includegraphics[width=60 mm]{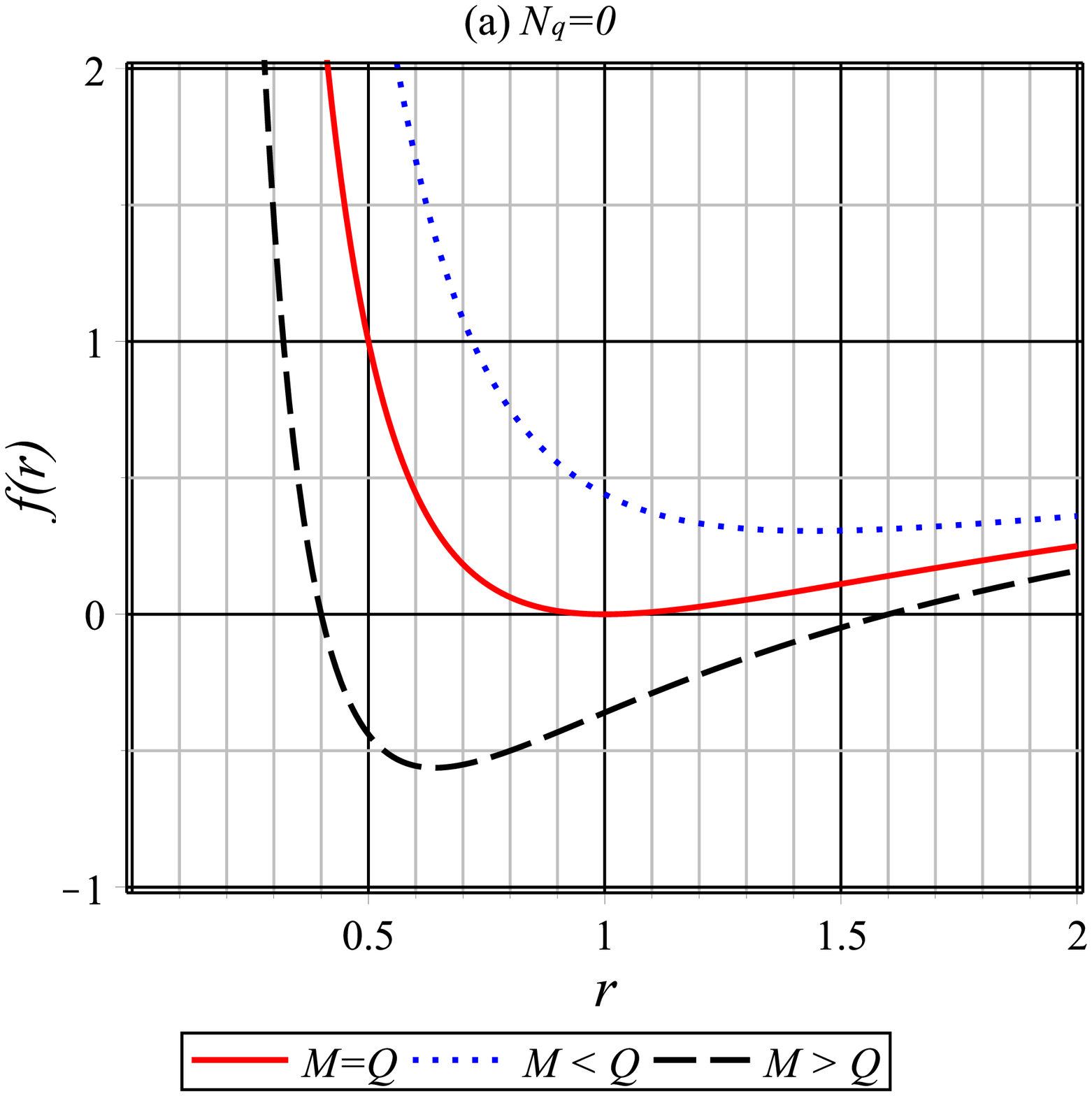}\includegraphics[width=60 mm]{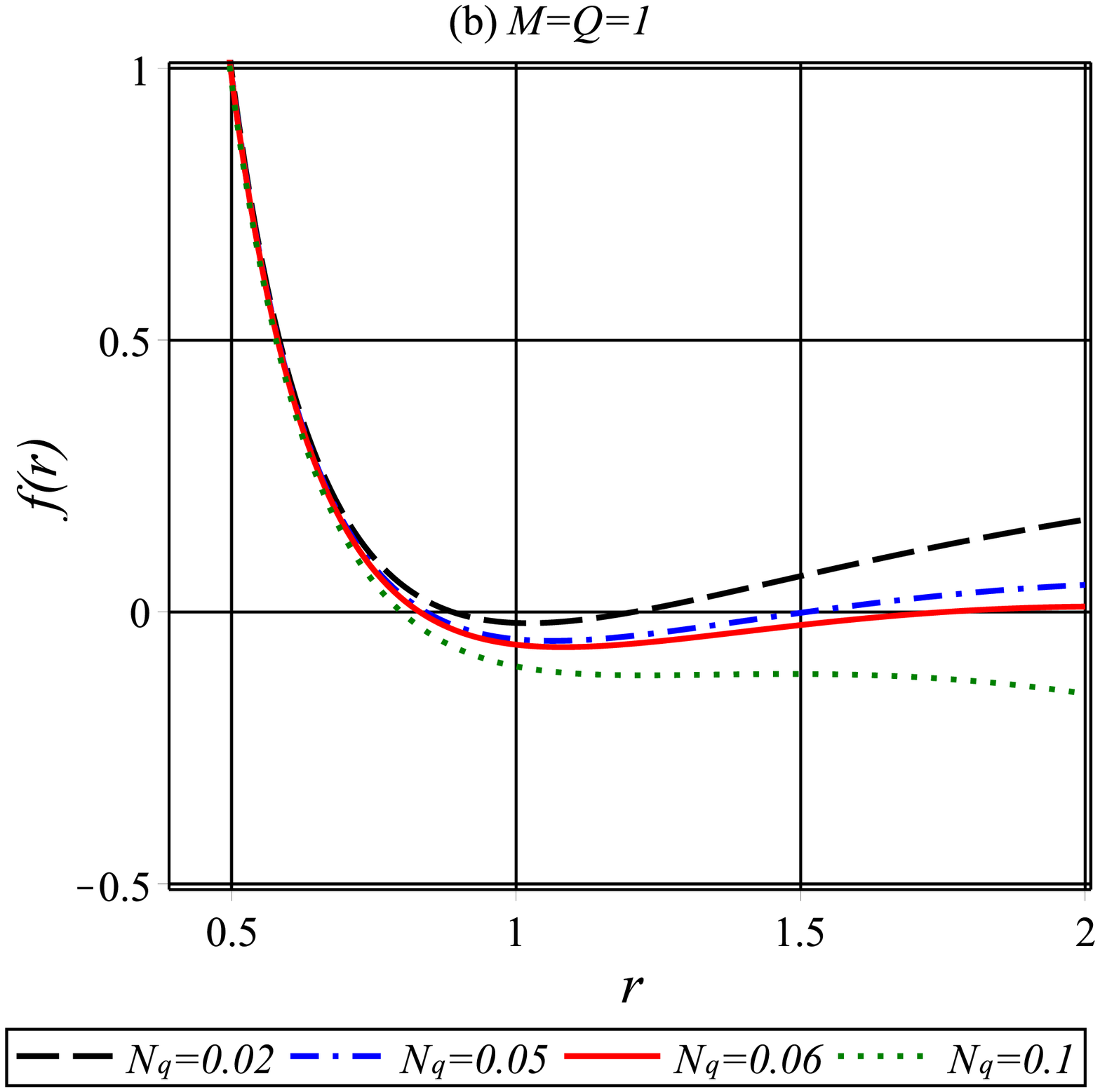}
 \end{array}$
 \end{center}
\caption{Horizon structure of the black hole surrounded by the quintessence field with $\omega_{q}=-\frac{2}{3}$ and $k\lambda=\frac{1}{4}$. (a) The case of $N_q=0$. Solid red line drawn for $M=Q=1$ which is extremal case. (b) Variation of quintessence
parameter $N_{q}$ for $M=Q=1$.}
 \label{fig1}
\end{figure}
In the plots of Fig. \ref{fig1},  we can see the horizon structure of the black hole surrounded by the quintessence field in Rastall theory. We can see extremal case corresponding to $N_{q}=0$ and $M=Q$, while there is naked singularity for the $Q>M$ and regular black hole for $M>Q$. It is possible to have $r_{+,q}$ and $r_{-,q}$, which is illustrated by the Fig. \ref{fig1}.

Hence, by using the relations (\ref{4}) and (\ref{6}), one can obtain the following relation for the black hole mass \cite{sud0}:
\begin{equation}\label{8}
M=\frac{\pi^{2}Q^{2}+\pi S-N_{q}S^{2}}{2\pi^{\frac{3}{2}}\sqrt{S}}.
\end{equation}
Consequently, the black hole temperature takes the following form:
\begin{equation}\label{9}
T=\frac{\pi S-3N_{q}S^{2}-\pi^{2}Q^{2}}{4\sqrt{\pi^{3}S^{3}}}.
\end{equation}
Also, the heat capacity is given by,
\begin{equation}\label{10}
C=\frac{2S(\pi S-3N_{q}S^{2}-\pi^{2}Q^{2})}{3\pi^{2}Q^{2}-\pi S-3N_{q}S^{2}}.
\end{equation}
In the Ref. \cite{sud0}, it has been argued that the stable-unstable phase transition exist.
\subsection{Dust field}
The black hole surrounded by the dust field is given by setting $\omega_{s}\equiv\omega_{q}=0$ \cite{HD}. Also, we assume $k\lambda=\frac{2}{9}$ and find that the equation (\ref{3}) reduces to the following relation:
\begin{equation}\label{15}
1-\frac{2M}{r}+\frac{Q^{2}}{r^{2}}-N_{d}r=0,
\end{equation}
where $N_s\equiv N_{d}$ is the dust field parameter and $\Lambda_{s}\equiv\Lambda_{d}=-1$. So, the black hole event horizon radius is given by 
\begin{equation}\label{16}
r_{+,d}=\frac{X^{\frac{1}{3}}}{6N_{d}}+\frac{2(1-6MN_{d})}{3N_{d}X^{\frac{1}{3}}}+\frac{1}{3N_{d}},
\end{equation}
where
\begin{equation}\label{17}
X=108\,{Q}^{2}{N}^{2}+12\,\sqrt {3}\sqrt {27\,{N}^{2}{Q}^{4}+32\,{M}^{3}
N-36\,MN{Q}^{2}-4\,{M}^{2}+4\,{Q}^{2}}N-72\,MN+8.
\end{equation}
Therefore, by using the relations (\ref{4}) and (\ref{6}) one can obtain the following relation for the black hole mass \cite{sud0}:
\begin{equation}\label{18}
M=\frac{\pi^{2}Q^{2}+\pi S-N_{d}\sqrt{\pi S^{3}}}{2\pi^{\frac{3}{2}}\sqrt{S}}.
\end{equation}
In the case of dust field, the black hole temperature is given by 
\begin{equation}\label{19}
T=\frac{\pi S-2N_{d}\sqrt{\pi S^{3}}-\pi^{2}Q^{2}}{4\sqrt{\pi^{3}S^{3}}}.
\end{equation}
the heat capacity is given by,
\begin{equation}\label{20}
C=\frac{2S(\pi S-2N_{d}\sqrt{\pi S^{3}}-\pi^{2}Q^{2})}{3\pi^{2}Q^{2}-\pi S}.
\end{equation}
This case also leads to the black hole phase transition.
\subsection{Radiation field}
The black hole surrounded by the radiation field is given by setting $\omega_{s}\equiv\omega_{r}=\frac{1}{3}$  \cite{1908.09629}.
Assuming $k\lambda=\frac{2}{9}$, the equation (\ref{3}) takes the following form:
\begin{equation} 
1-\frac{2M}{r}+\frac{Q^{2}}{r^{2}}-\frac{N_{r}}{r^{2}}
=0,
\end{equation}
where $N_s\equiv N_{r}$ is the radiation field parameter and $\Lambda_{s}\equiv\Lambda_{r}=2$. Hence, we   obtain 
\begin{eqnarray}\label{2323}
r_{\pm,r}=M\pm\sqrt{M^{2}-Q^{2}+N_{r}}.
\end{eqnarray}
It means that the extremal case is given by $M^{2}=Q^{2}-N_{r}$, while naked singularity occurs for $M^{2}+N_{r}<Q^{2}$.
Therefore, by using the relations (\ref{4}) and (\ref{6}), we obtain the following expression for the black hole mass corresponding to radiation field:
\begin{eqnarray}
M=\frac{S+\pi Q^2-\pi N_r}{2\sqrt{\pi S}}.
\end{eqnarray}
In this case, the Hawking temperature is calculated by
\begin{eqnarray}
T=\frac{S-\pi Q^2+3\pi N_r}{4\sqrt{\pi S^3}}.
\end{eqnarray}
The heat capacity in this case is obtained as
\begin{eqnarray}
C=\frac{2S(S-\pi Q^2 +3\pi N_r)}{\pi Q^2-S-9\pi N_r}.
\end{eqnarray}
Here, it is clear from the condition (\ref{con}) that the first law of thermodynamics satisfied if $Q^{2}=N_{r}+c$.

\section{Thermal fluctuations}
In this section, we discuss how thermal fluctuations  around the thermal 
equilibrium affect the entropy of the black holes
 surrounded by perfect fluid in Rastall gravity.
 The  density of states in terms of the partition function $Z(\beta)$ 
 describing a black hole is written by
      \begin{equation}
       \rho{(E)} =\frac{1}{2\pi{i}}{  \int_{{\beta}_0 - i\infty}^{{\beta}_0+ 
       i\infty}{{d\beta}Z(\beta)e^{\beta{E}}}}
        = \frac{1}{2\pi{i}}{  \int_{{\beta}_0 - i\infty}^{{\beta}_0+ i\infty}
        {d\beta }e^{{\mathcal{S}}(\beta)}}.     \label{l}
        \end{equation}
where $\beta =  \frac{1}{T} $ since Boltzmann constant is unit here and $S = 
\ln{Z(\beta)} + \beta{E} $ is exact entropy for the  black hole. For small 
sized   black hole, we  expand entropy  around equilibrium and apply method  
of steepest descent (where $\frac{d  S}{d\beta} = 0$ and $\frac{d^2  S}
{d{\beta}^2}  >  {0}$) to get
      \begin{equation}\label{eqp}
       {\mathcal{S}}(\beta) = S + \frac{1}{2}({\beta} - {\beta}_0)^2\frac{d^2  
       S}{d{\beta}^2}|
       _{{\beta} = {\beta}_0}  + \mbox{(higher   order  terms)},
      \end{equation}
      where $S$ represents   equilibrium    entropy.
  By  utilizing relations (\ref{eqp}) and (\ref{l}),   we have
            \begin{equation}
       \rho{(E)} = \frac{e^{S}}{\sqrt{2\pi{\frac{d^2  S}{d{\beta}^2}}}}.
      \end{equation}
       Eventually, the logarithm
of this density of states leads to the microcanonical entropy
\begin{equation}\label{main}
{\mathcal{S}} = S -\frac{1}{2}\ln\left(\frac{d^2  S}{d{\beta}^2}\right).
\end{equation}
It is usual to add a constant coefficient to the logarithmic terms to track correction terns \cite{EPL}. But here, because our main results are based on numerical analysis, we don't require such constant coefficient.
Next, we examine the explicit values of entropy for different fields.
\subsection{Quintessence field}
In this case, the entropy (\ref{main}) reduces to the following expression:
\begin{equation}\label{mainq}
{\mathcal{S}}=\pi (r_{+,q})^{2}-\frac{1}{2}\ln\left({\frac { \left(3N_{q}({r_{+,q}})^{4}-({r_{+,q}})^{2}+{Q}^{2}\right)^{3}}{54\pi({r_{+,q}})^{6} \left(N_{q}({r_{+,q}})^{4}+\frac{({r_{+,q}})^{2}}{3} -{Q}^{2}\right) ^{3}}}X_{q}\right),
\end{equation}
where $X_{q}$ is defined by
\begin{equation}
X_{q}=9{N_{q}}^{2}({r_{+,q}})^{8}+6N_{q} ({r_{+,q}})^{6}-({r_{+,q}})^{4}(1+30N_{q}{Q}^{2})+6{Q}^{2}({r_{+,q}})^{2}-3{Q}^{4}.
\end{equation}

\subsection{Dust field}
For the case of dust field,  the entropy (\ref{main}) takes the following relation:
\begin{equation}\label{maind}
{\mathcal{S}}=\pi (r_{+,d})^{2}-\frac{1}{2}\ln\left({\frac{\left(2N\sqrt{\pi}({r_{+,d}})^{3}-({r_{+,d}})^{2}+{Q}^{2}\right)^{3}}{2\pi^{\frac{15}{2}}({r_{+,d}})^{13} \left(3{Q}^{2} -({r_{+,d}})^{2}\right) ^{3}}}X_{d}\right),
\end{equation}
where $X_{d}$ is defined as
\begin{equation}
X_{d}=15N{Q}^{2}{\pi }^{7}({r_{+,d}})^{12}-3N{\pi }^{7}({r_{+,d}})^{14}+\sqrt{{\pi}^{13}}\left( 3{Q}^{4}-6{Q}^{2}({r_{+,d}})^{2}+({r_{+,d}})^{4} \right) ({r_{+,d}})^{9}.
 \end{equation}
\subsection{Radiation field}
\begin{figure}[h!]
 \begin{center}$
 \begin{array}{cccc}
\includegraphics[width=50 mm]{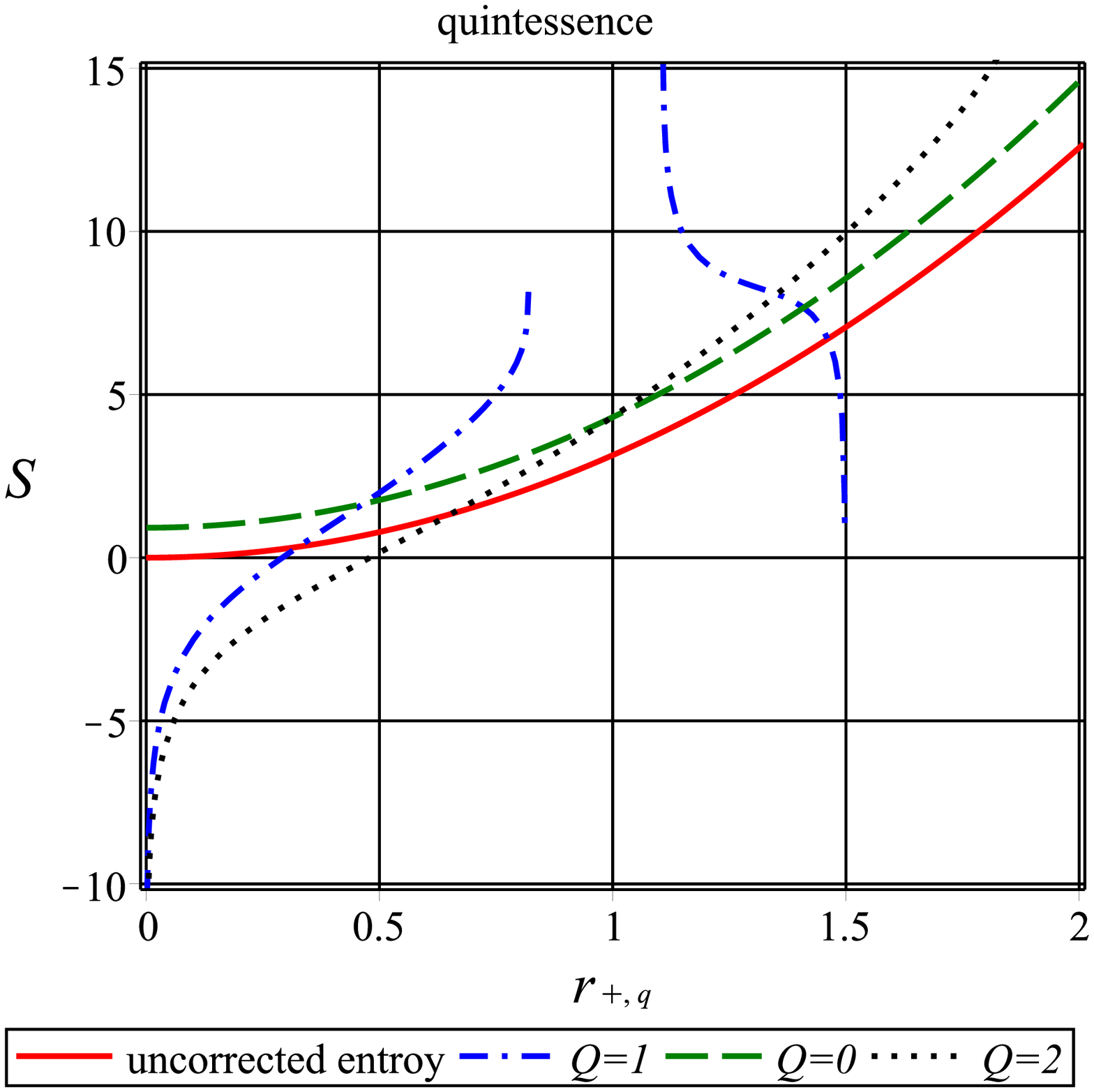}&\includegraphics[width=50 mm]{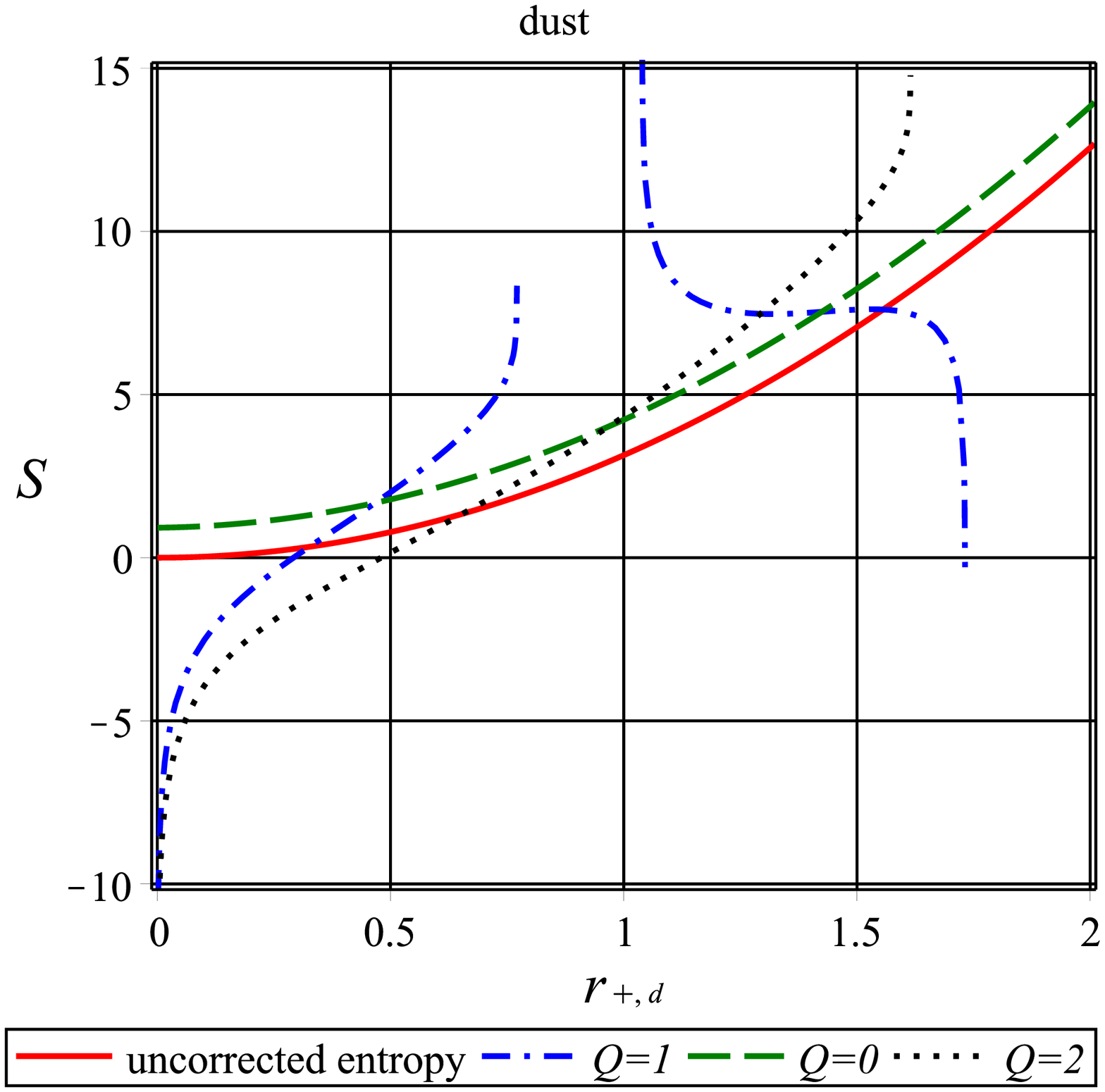}&\includegraphics[width=50 mm]{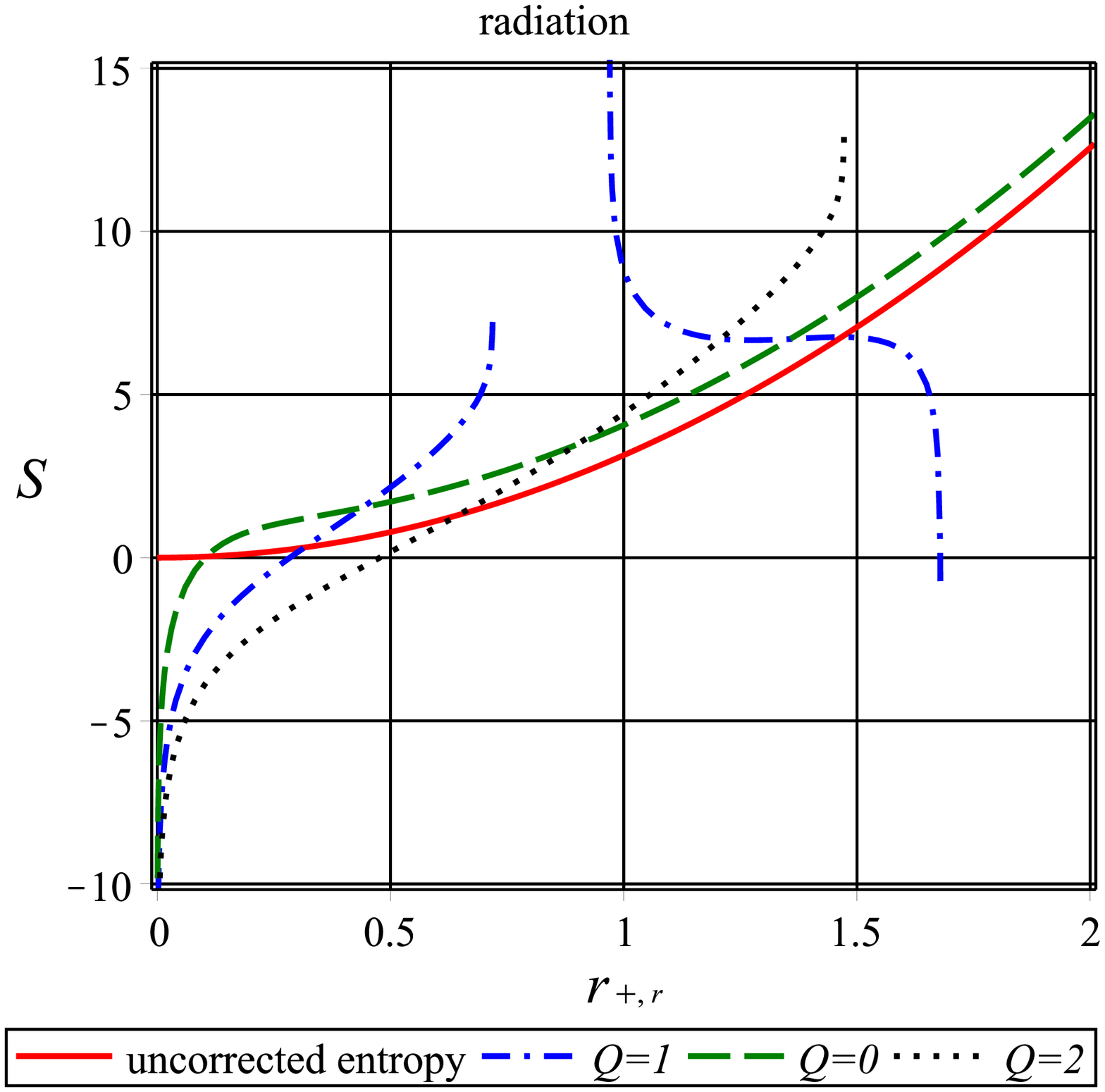}
 \end{array}$
 \end{center}
\caption{Typical behaviour of the entropy in terms of horizon radius, for $N_{s}=0.02$.}
 \label{fig2}
\end{figure}

\begin{figure}[h!]
 \begin{center}$
 \begin{array}{cccc}
\includegraphics[width=65 mm]{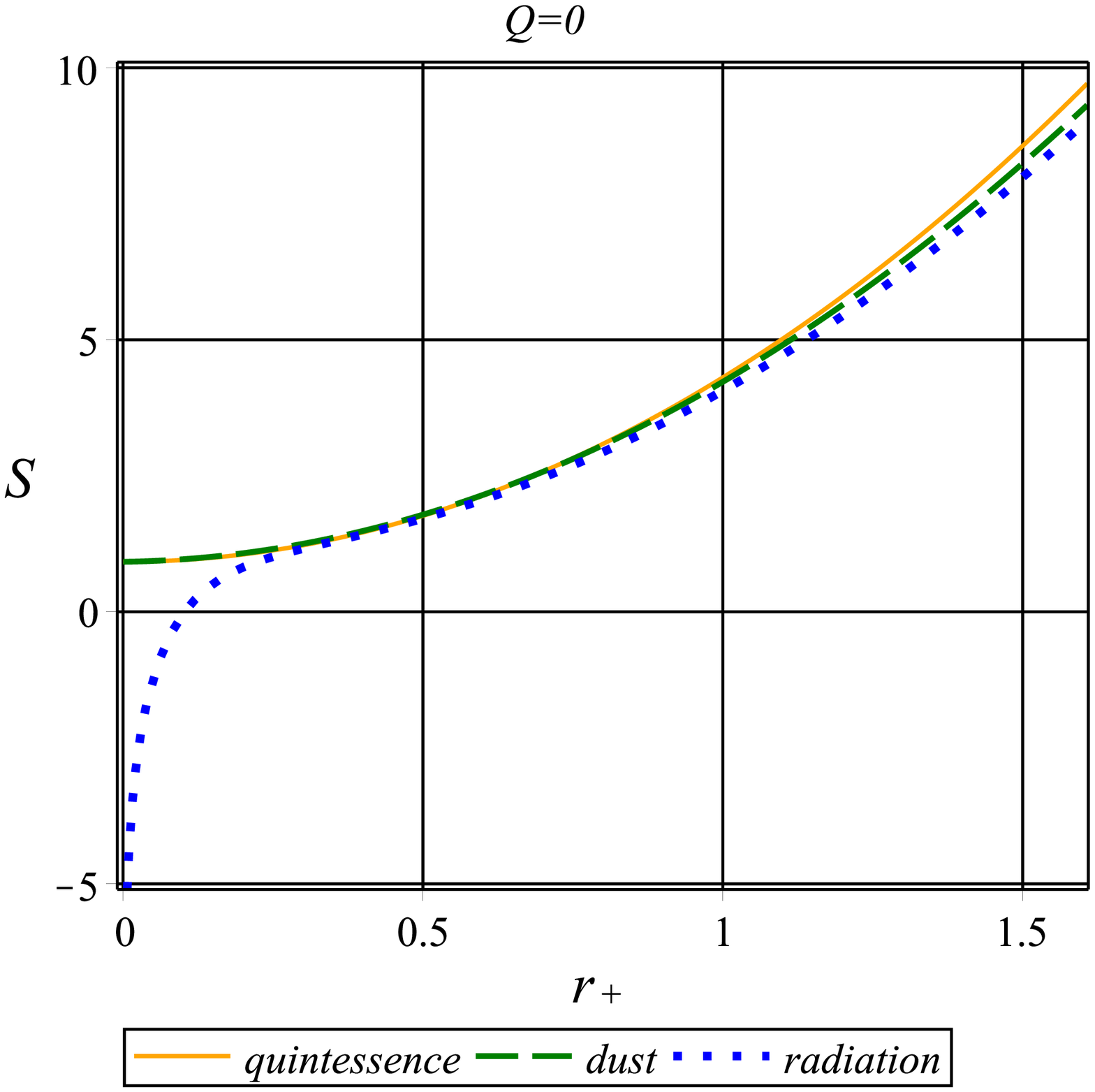}&\includegraphics[width=65 mm]{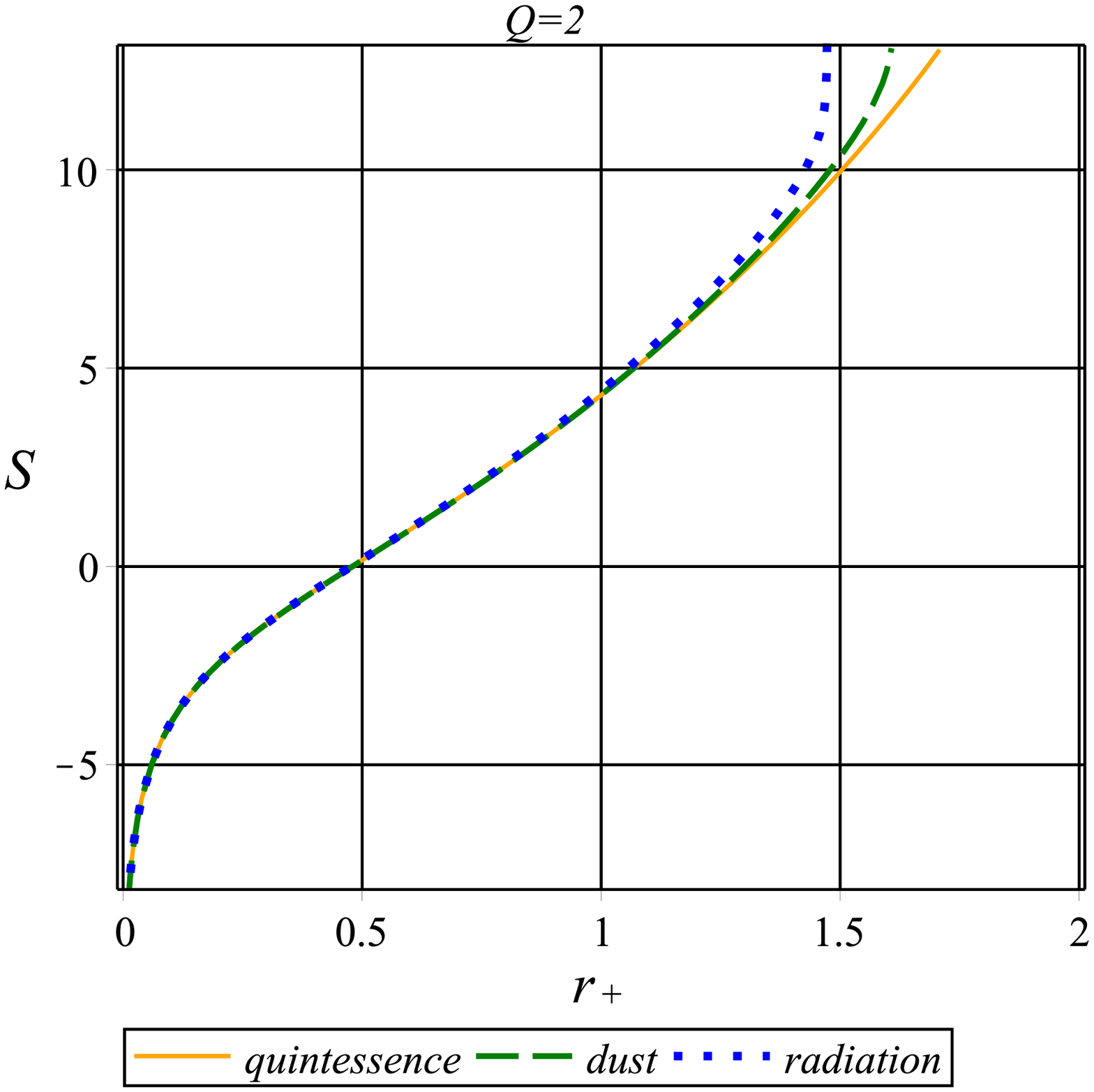}
 \end{array}$
 \end{center}
\caption{Typical behaviour of the corrected entropy in terms of horizon radius  for $N_{s}=0.02$.}
 \label{fig3}
\end{figure}

In the  case of radiation field, the entropy (\ref{main}) reduces to the following equation:
\begin{equation}
{\mathcal{S}}=\pi (r_{+,r})^{2}-\frac{1}{2}\ln\left({\frac{27\left(3N+({r_{+,r}})^{2}-{Q}^{2}\right)^{3}}{2\pi^{3}({r_{+,r}})^{4} \left(9N+({r_{+,r}})^{2}-3{Q}^{2}\right) ^{3}}}X_{r}\right),
\end{equation}
where we $X_{r}$ is defined as
\begin{equation}
X_{r}=\frac{{\pi}^{2}}{9}\left(\frac{({r_{+,r}})^{4}}{3}+(3N-Q^{2})((2{r_{+,r}})^{2}+3N-Q^{2})\right).
 \end{equation}
In the plots of the Fig. \ref{fig2}, we draw entropy in terms of horizon radius for three cases of quintessence, dust and radiation fields. The first plot shows the entropy of black hole surrounded by quintessence field in Rastall theory. We can see that corrected entropy may increase or decrease. Also, corrected entropy is negative for small radius. It means that in quantum level the black hole may be unstable. Later, we discuss about the black hole stability by analysing specific heat. We can see approximately similar behaviour for the black hole surrounded by dust and radiation fields in Rastall theory which is illustrated by the second and third plots. In all 
the cases, we can see an asymptotic behaviour (dash dotted blue lines which is corresponding to $Q=1$) which may be sign of phase transition. It should be verified by analysing specific heat.
In order to compare three different cases of quintessence, dust and radiation entropy with each other we put them in single plot for $Q=0$ and other plot for $Q=2$ (see Fig. \ref{fig3}). Right plot of the Fig. \ref{fig3} shows that the charged system has the same behaviour at small radius. In the case of $Q=0$ (left plot of the Fig. \ref{fig3}) the black hole surrounded by radiation field yields to negative entropy at small radii which may be sign of instability at quantum level. It means that   thermal fluctuations 
increases instability of black hole.
\section{Thermodynamics}
In order to calculate thermodynamical variables, we fix black hole volume (horizon radius). Having entropy and temperature, one can calculate other thermodynamics quantities like Helmholtz free energy defined as
\begin{equation}\label{F}
F=-\int{{\mathcal{S}} dT}.
\end{equation}
For a given Helmholtz free energy, the internal energy is given by 
\begin{equation}
U=F+T{\mathcal{S}}.
\end{equation}
Using internal energy, we can obtain specific heat at constant volume
from the relation,
\begin{equation}\label{C}
C_{V}=\left(\frac{\partial U}{\partial T}\right)_{V}.
\end{equation}
The analysis of its sign we study the black hole stability and its asymptotic behaviour tells us about the black hole phase transition.

The study of another interesting characteristics left  is behaviour of $p-V$ diagram. By using the black hole volume 
\begin{equation}\label{V}
V=\frac{4}{3}\pi r_{+}^{3},
\end{equation}
one can calculate black hole pressure via the following relation:
\begin{equation}\label{p}
p=-\left(\frac{\partial F}{\partial V}\right)_{T}.
\end{equation}
Therefore, by drawing a plot of $p$ in terms of $V$ we can study the critical points.

Once the internal energy, pressure and black hole volume are known, one can obtain the enthalpy as
\begin{equation}
H=U+pV,
\end{equation}
which finally leads to the Gibbs free energy as following:
\begin{equation}
G=H-T{\mathcal{S}}.
\end{equation}
Now, we shall implement  above points for three different cases of black hole surrounded by quintessence, dust and radiation fields in the forthcoming subsections.
\subsection{Quintessence field}
\begin{figure}[h!]
 \begin{center}$
 \begin{array}{cccc}
\includegraphics[width=50 mm]{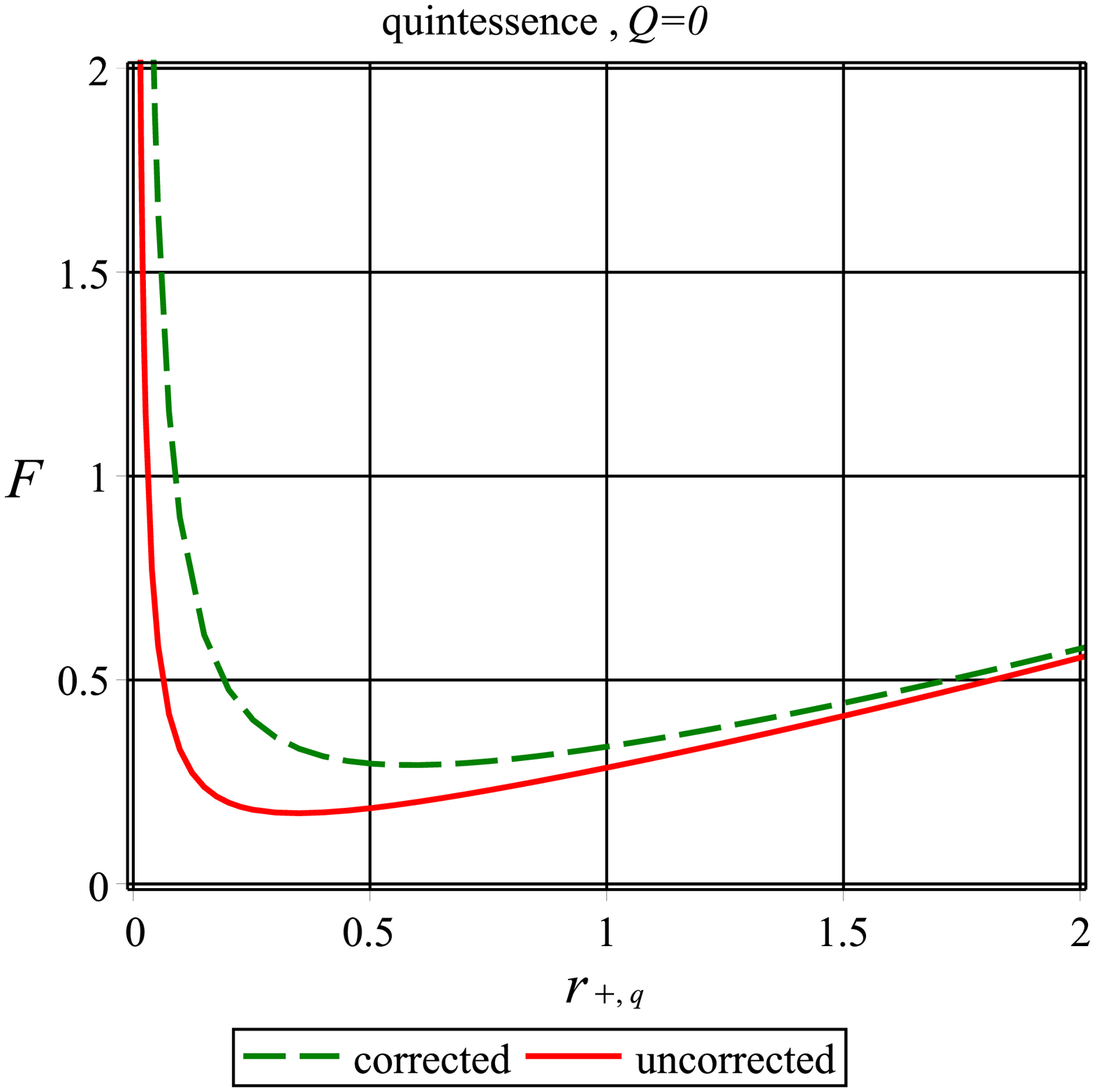}&\includegraphics[width=50 mm]{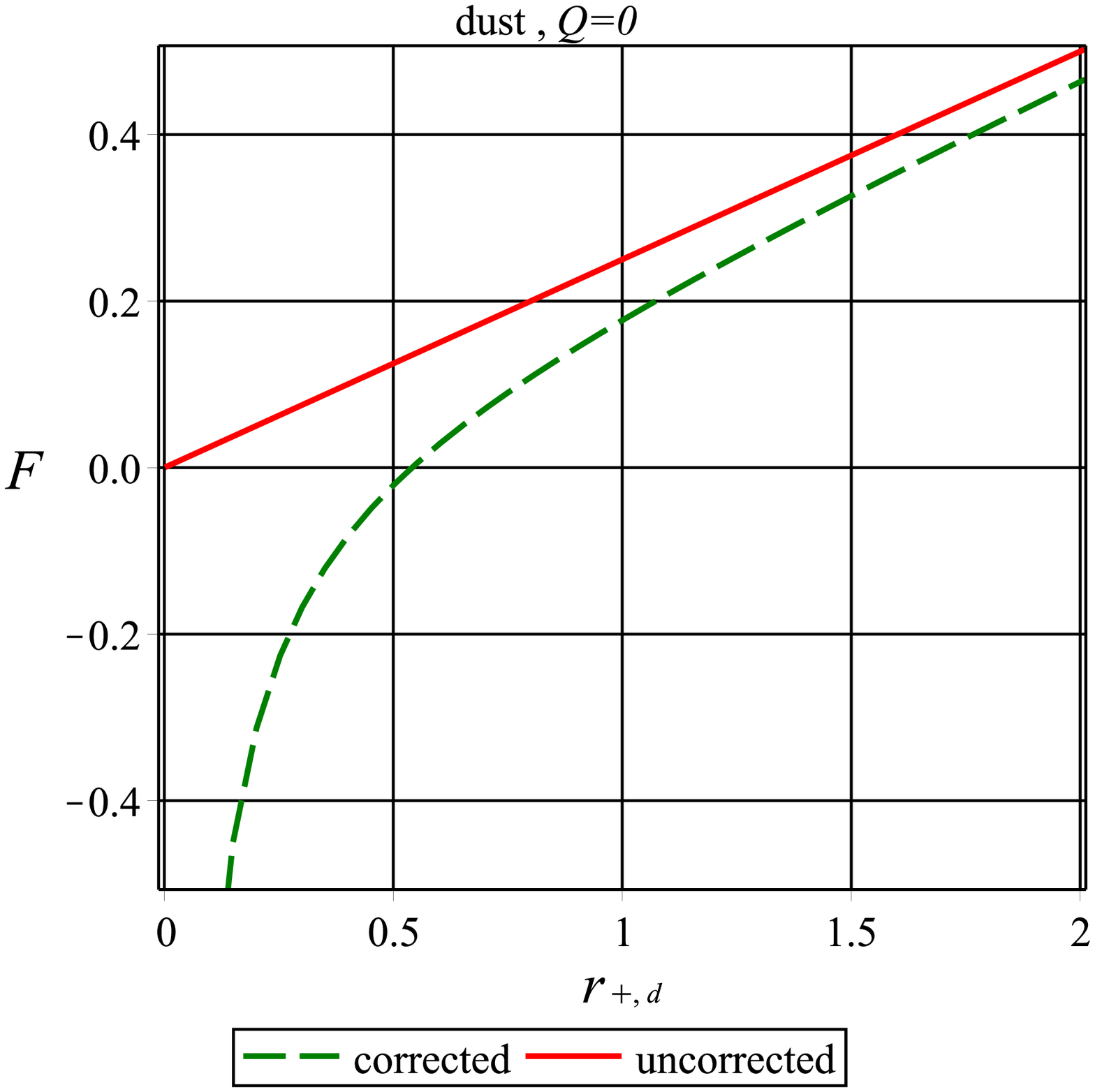}
&\includegraphics[width=50 mm]{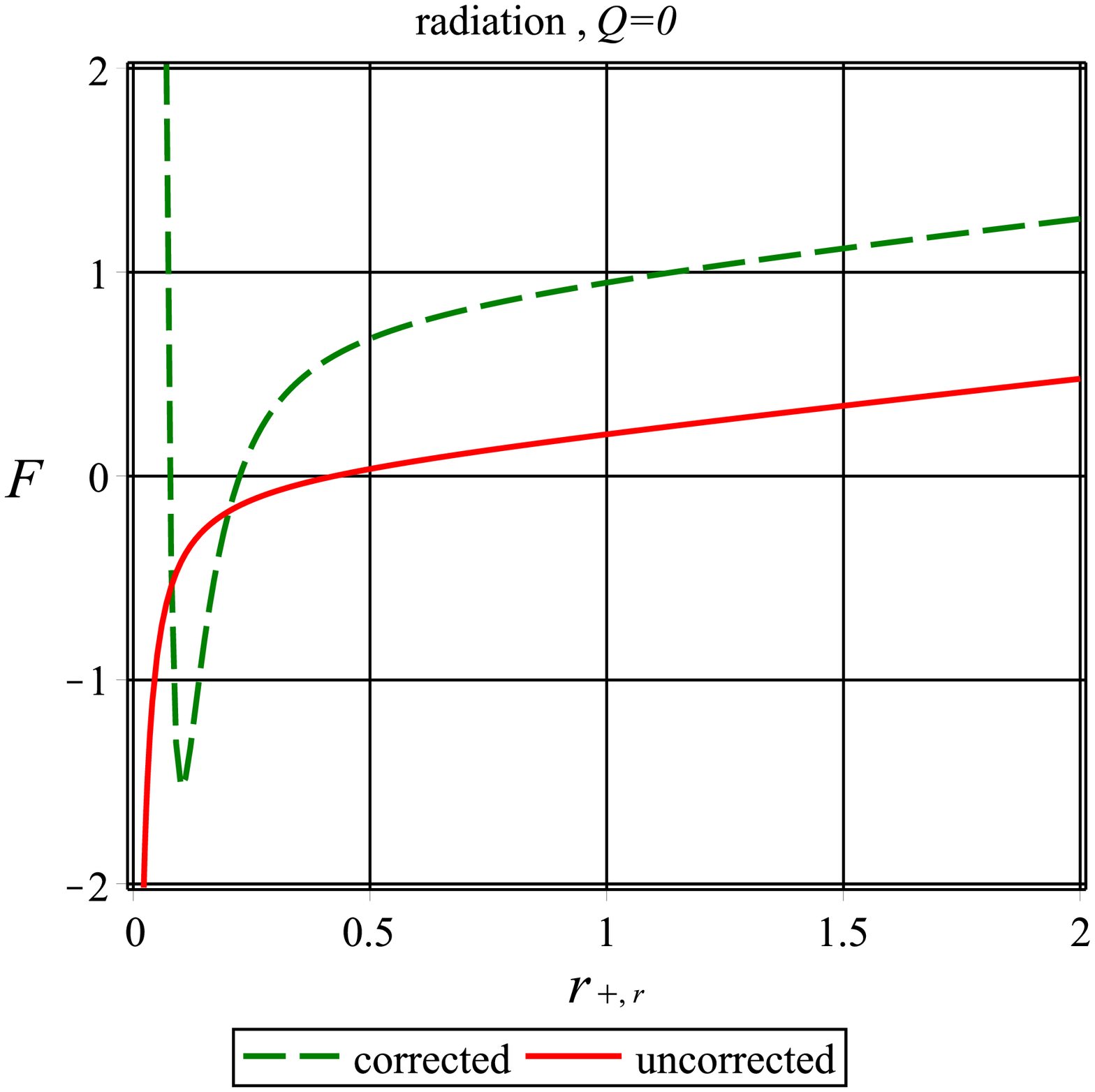}\\
\includegraphics[width=50 mm]{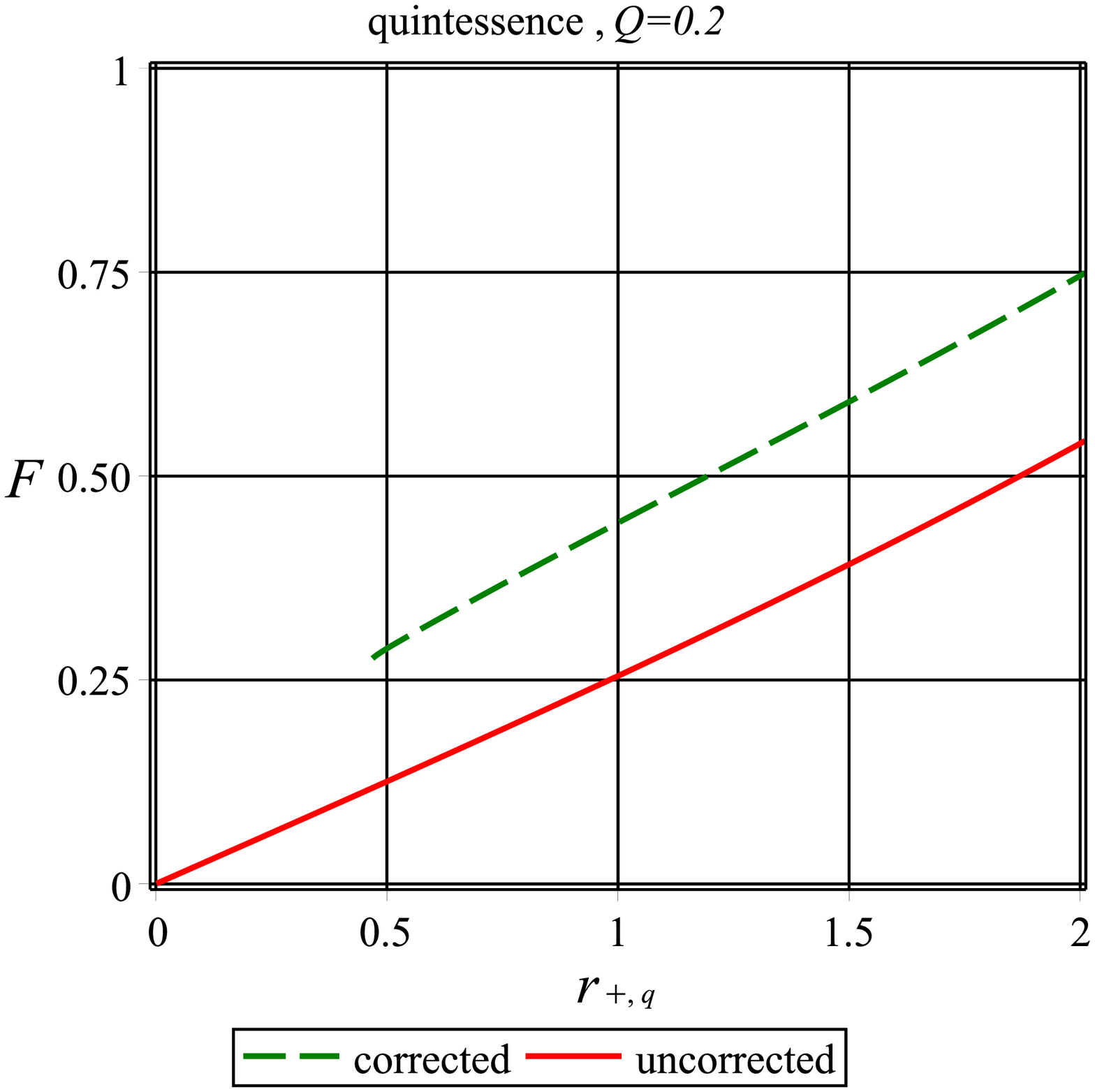}&\includegraphics[width=50 mm]
{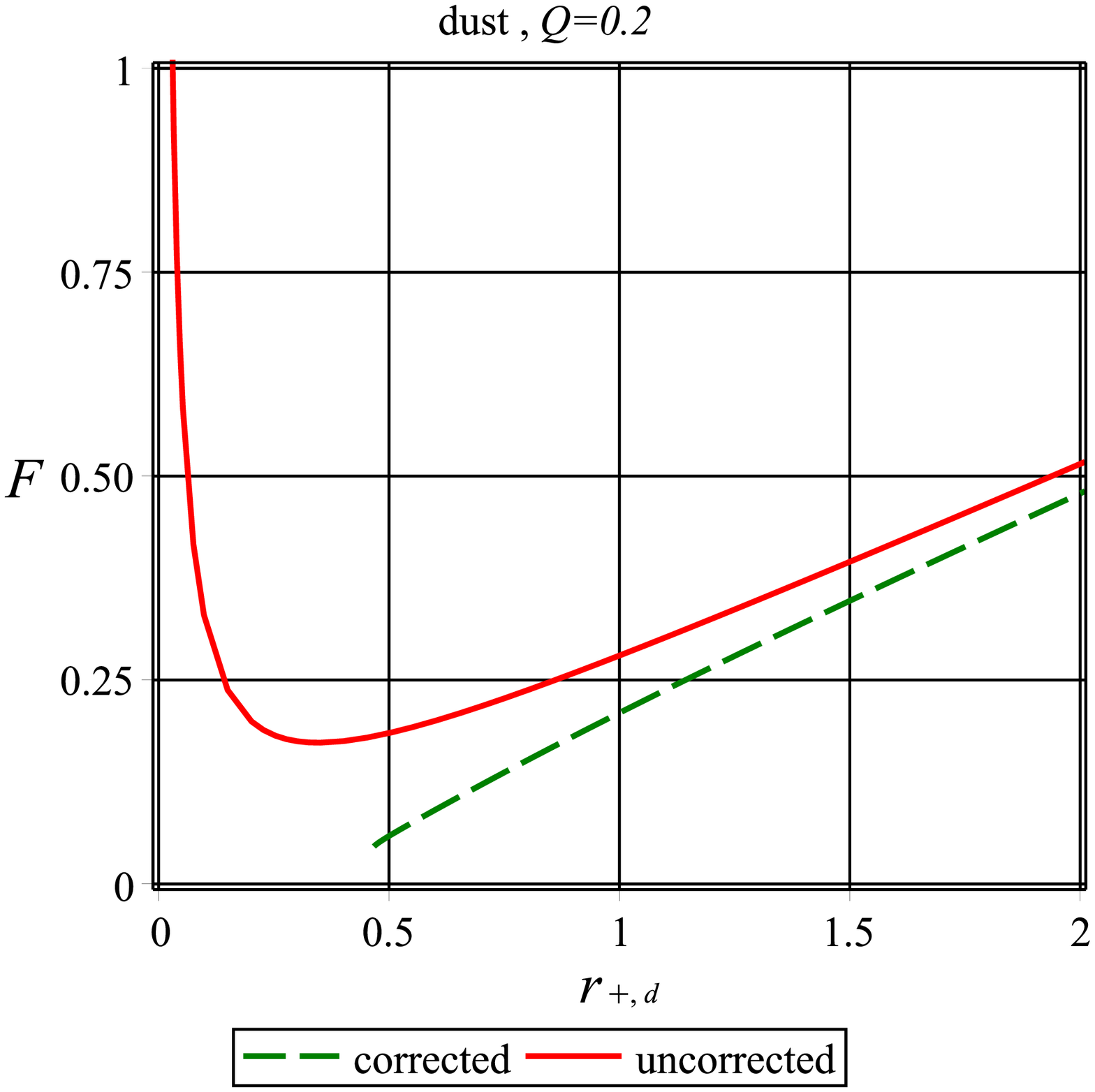}&\includegraphics[width=50 mm]{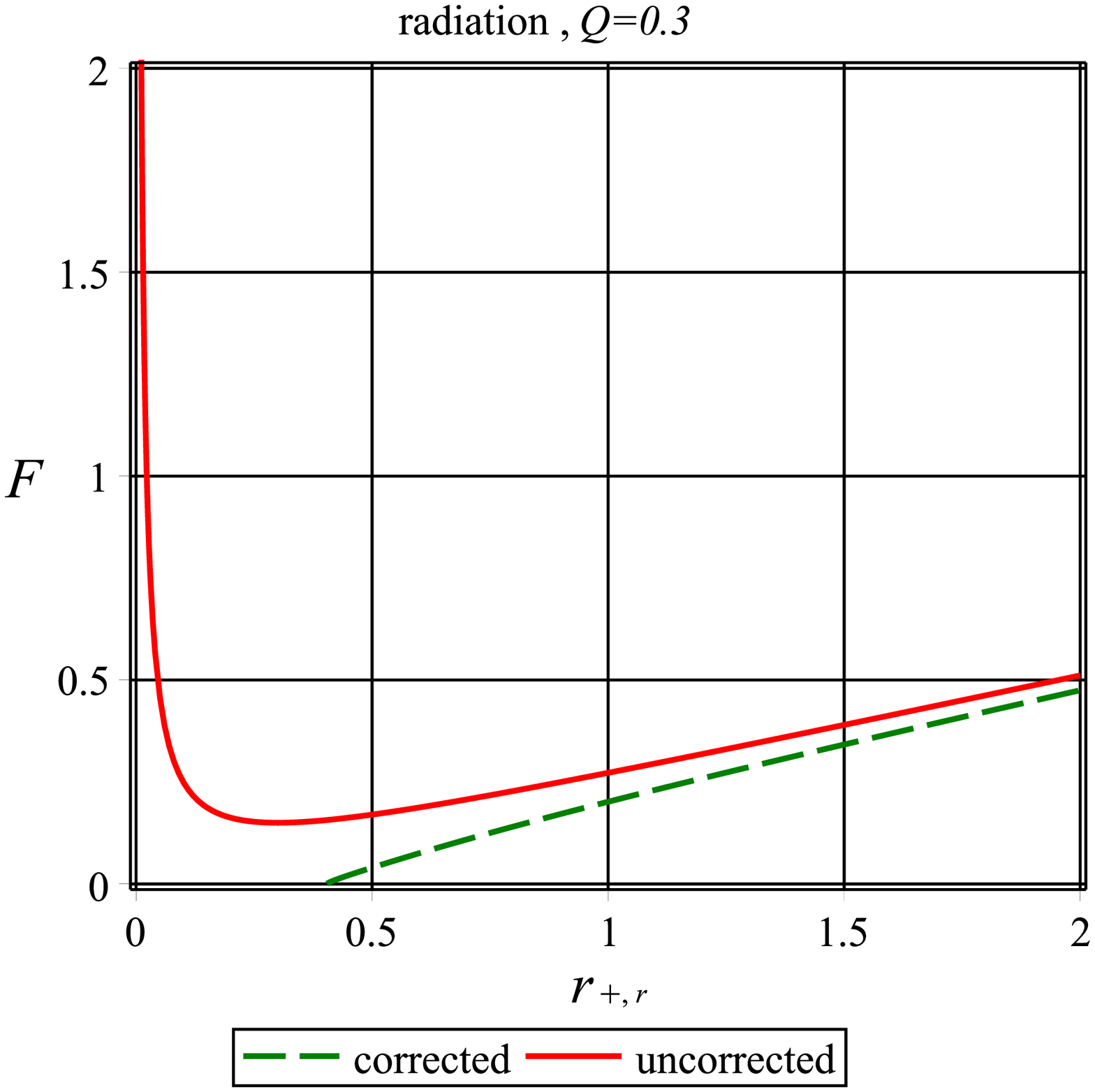}
 \end{array}$
 \end{center}
\caption{Typical behaviour of the Helmholtz free energy in terms of horizon 
radius  for $N_{q}=0.02$ (left plots), $N_{d}=0$ (middle plots) and $N_{r}
=0.02$ (right plots).}
 \label{fig4}
\end{figure}

Neglecting the thermal fluctuations and  using the equation (\ref{F}), we can 
write the expression of  Helmholtz free energy for the case of quintessence 
field as follows,
\begin{equation}
F={\frac {({r_{+,q}})^{4}N_{q}+({r_{+,q}})^{2}+3{Q}^{2}}{4 {r_{+,q}} }}.
\end{equation}
In presence of thermal fluctuation, we find that Helmholtz free energy 
increases. In some cases of charged black hole surrounded by quintessence 
field in Rastall theory with thermal fluctuation, there is a minimum bound 
for the horizon radius (see left plots of the Fig. \ref{fig4}) which may be 
sign of some instabilities which is mentioned already.

\subsection{Dust field}
Neglecting the thermal fluctuations and making use the equation (\ref{F}), 
we   obtain Helmholtz free energy of black hole surrounded by dust field 
in Rastall theory as follow,
\begin{equation}
F={\frac {({r_{+,d}})^{2}+3{Q}^{2}}{4 {r_{+,d}} }}.
\end{equation}
Surprisingly, it is independent of $N_{d}$, hence we draw behaviour of the 
Helmholtz free energy in presence of thermal fluctuations for the case of 
$N_{d}=0$ to compare with uncorrected case (see middle plots of the Fig. 
\ref{fig4}). Also, we can see opposite behaviour with the previous case as
the thermal fluctuation decreases the Helmholtz free energy. 

Other thermodynamics potentials like enthalpy, internal and Gibbs free energy 
have similar behaviour.

\subsection{Radiation field}

Neglecting the thermal fluctuations and exploiting equation (\ref{F}), we calculate Helmholtz free energy for black hole surrounded by radiation field in 
Rastall theory as follows,
\begin{equation}
F={\frac {({r_{+,r}})^{2}+3{Q}^{2} -9N_{r}}{4 {r_{+,r}} }}.
\end{equation}
Here, we find that Helmholtz free energy is zero if ${r_{+,r}}=
\sqrt{9N_{r}-3{Q}^{2}}$. Combining this with the equation (\ref{2323}), we 
obtain a condition on the black hole mass 
\begin{equation}
M^{4}-[2(4N_{r}-Q^{2})+1]M^{2}+Q^{2}-N_{r}=0.
\end{equation}
Root of this equation gives us a particular value of $M$ for which Helmholtz free energy is 
zero.

Effect of thermal fluctuations of this case is illustrated by right plots of the 
Fig. \ref{fig4}.
Here, we consider $Q=0.3$, because for the cases of $Q<0.3$ the 
situation is very similar to the uncharged black hole which is drawn in upper 
right plot. Such behaviour of free energy (a minimum with negative value) may 
be sign of some stabilities which will be verified in the next section. In the case 
of $Q=0$ we can see a new behaviour. Thermal fluctuations, depend on horizon 
radius, may reduce or increase value of Helmholtz free energy. Charged black 
hole with large value of charge behaves like previous case. 
\begin{figure}[h!]
 \begin{center}$
 \begin{array}{cccc}
\includegraphics[width=50 mm]{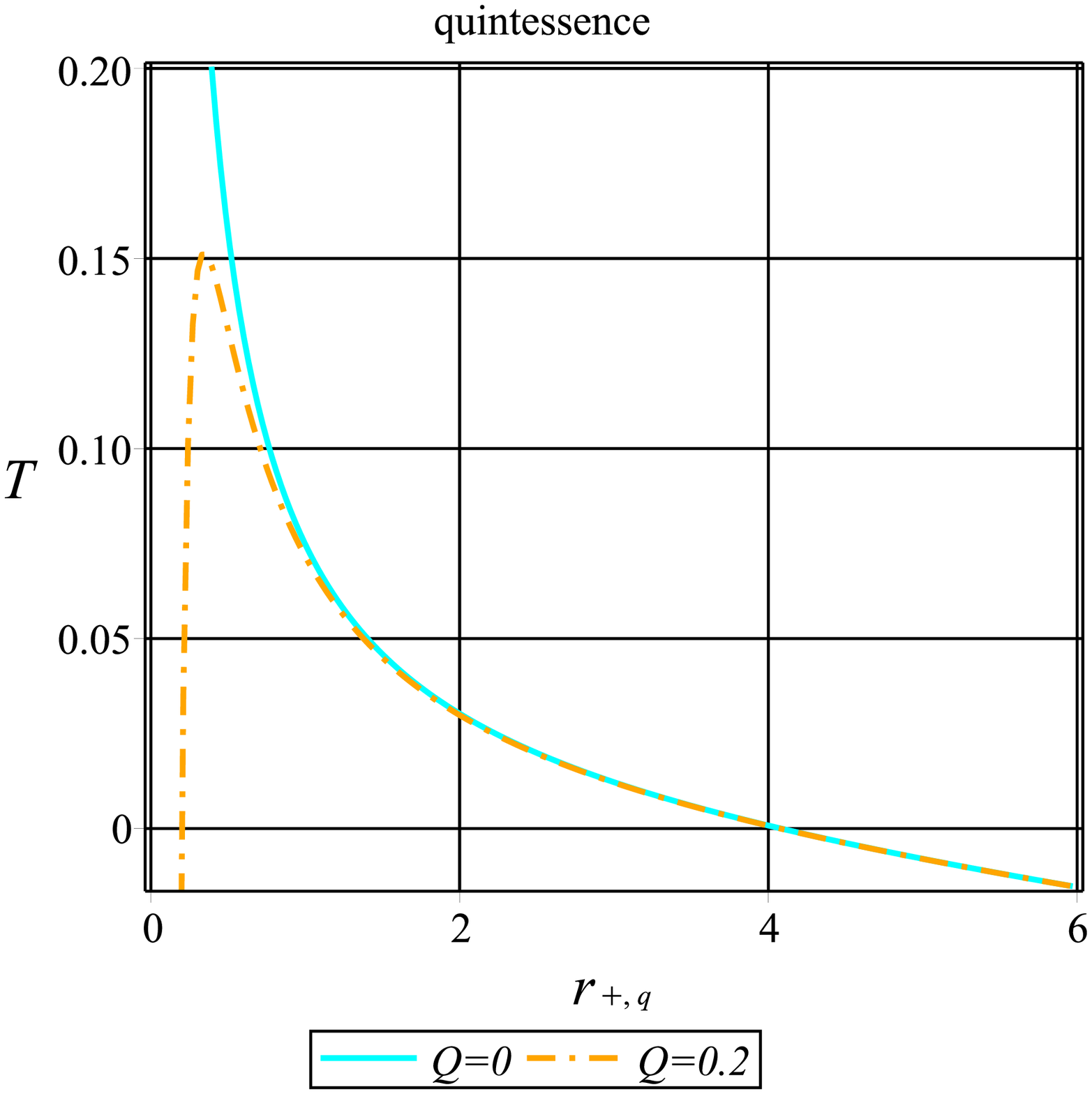}&\includegraphics[width=50 mm]{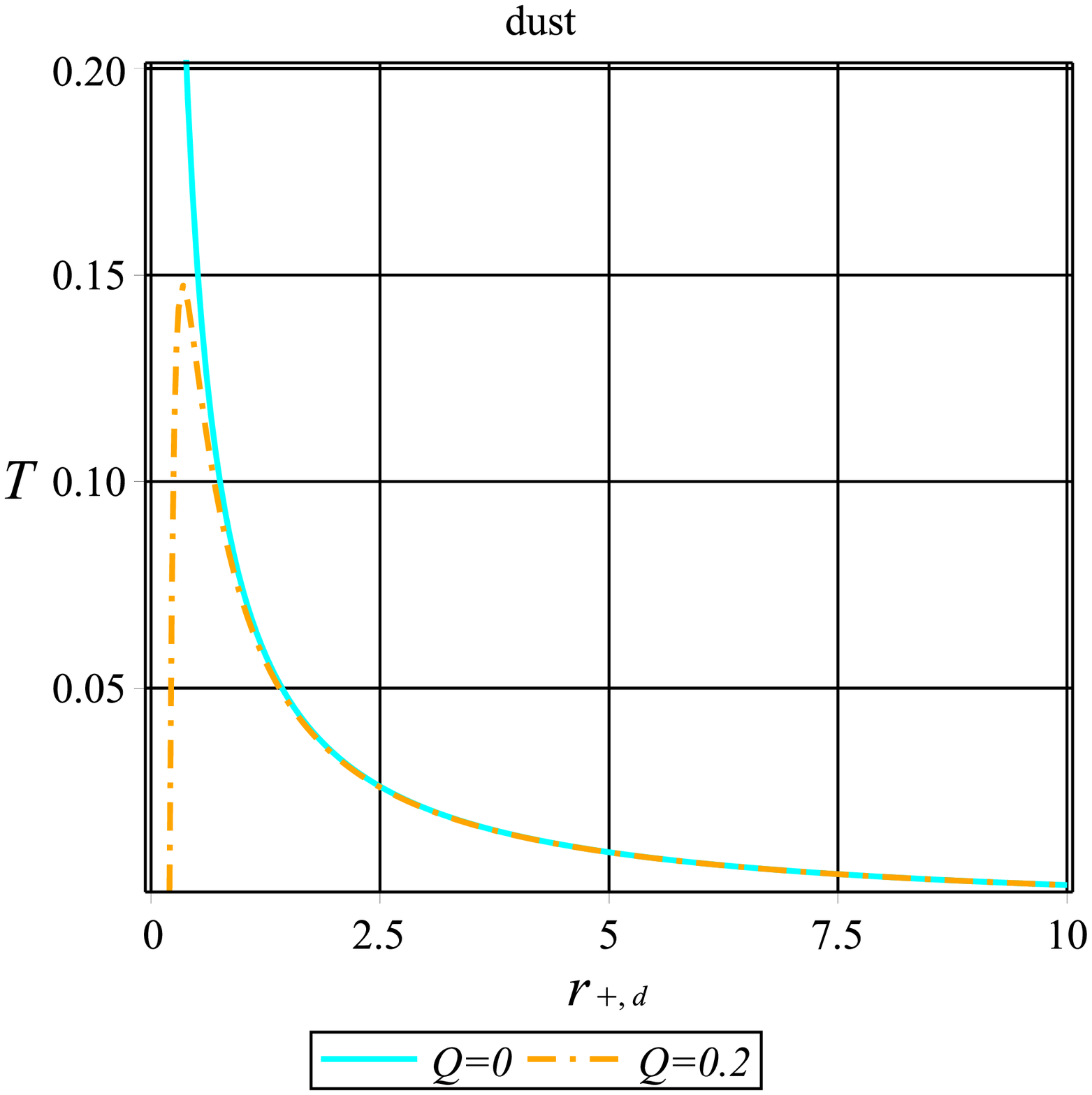}
&\includegraphics[width=50 mm]{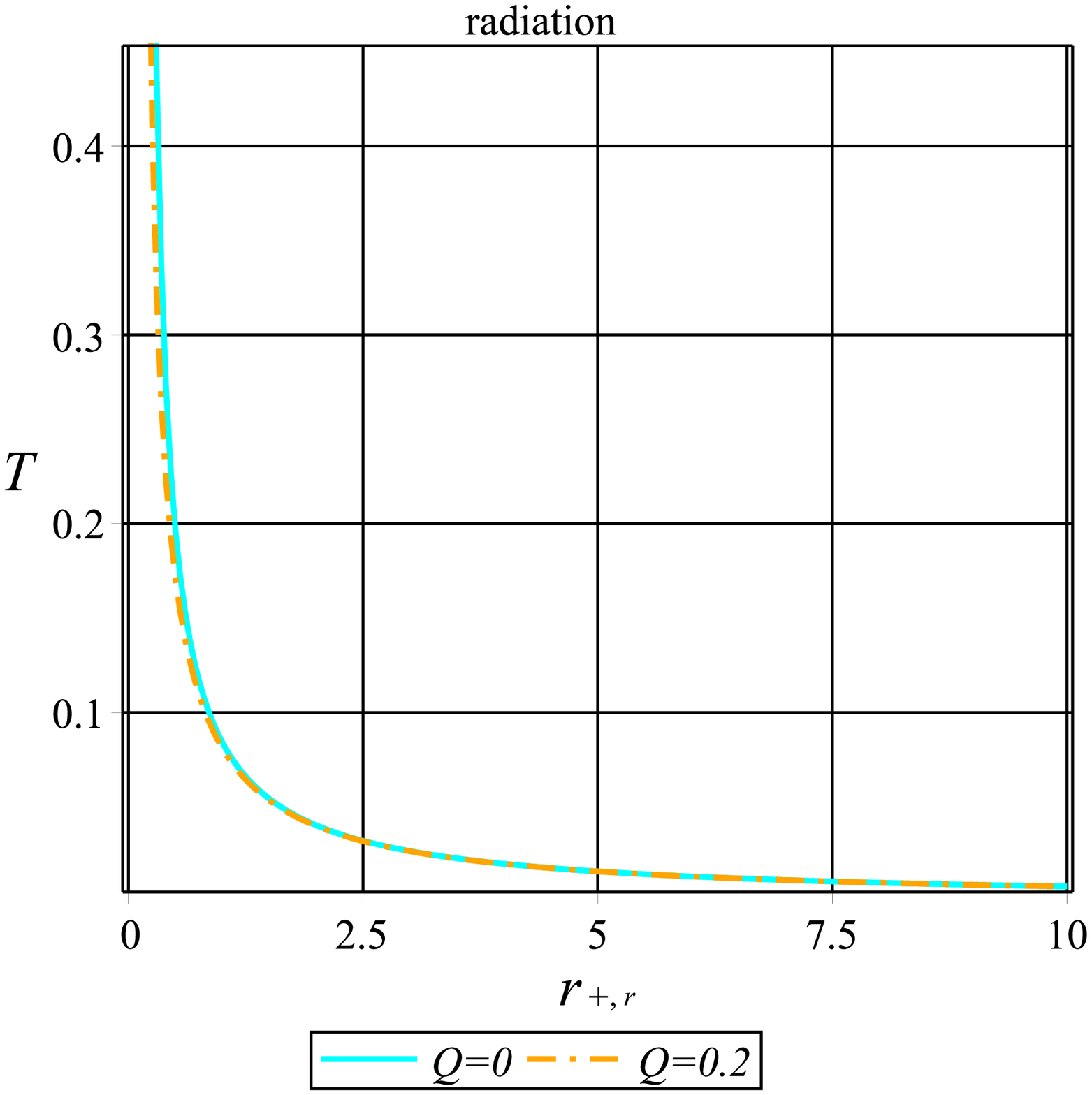}
 \end{array}$
 \end{center}
\caption{Typical behaviour of the temperature in terms of horizon radius, for 
$N_{s}=0.02$.}
 \label{fig5}
\end{figure}
In order to find a valid regions of horizon radius we draw plots of 
temperature 
with the fact that this must be positive. In the plots of the Fig. \ref{fig5}, we 
can see that temperature of black holes surrounded by dust and radiation 
fields are positive for $r_{+}<10$, hence for the selected values of our 
parameters there is no invalid regions, while for the black hole surrounded 
by quintessence field (left plot of the Fig. \ref{fig5})  we can see that 
temperature is negative for $r_{+}\geq4$, hence we must focus to this 
region.

\section{Critical points and stability}
In order to study black hole stability we should look at the specific heat at 
constant volume (\ref{C}). If $C_V>0$ then black hole is stable  and if $C_V<0$  then 
black hole is unstable, while asymptotic behaviour of the specific heat shows 
phase transition. We can also check the following conditions \cite{PV}:
\begin{eqnarray}\label{critical}
\frac{\partial p}{\partial V}&=&0,\nonumber\\
\frac{\partial^{2} p}{\partial V^{2}}&=&0.
\end{eqnarray}
These relations hold at critical points which are inflection points.

\subsection{Quintessence field}
Now, in the case of quintessence field,  the specific heat at constant volume is 
calculated by the equation (\ref{C}), which results to the following expression:
\begin{equation}\label{CQ}
C_{V}=\frac{1}{\left( 3({r_{+,q}})^{4}N_{q}+({r_{+,q}})^{2}-3{Q}^{2} 
\right)^{2}}\frac{A_{q}}{B_{q}},
\end{equation}
where $A_{q}$ and $B_{q}$  are defined as
\begin{eqnarray}
A_{q}&=&162{N_{q}}^{4}\pi ({r_{+,q}})^{18}+54 \left(2\pi-3{N_{q}} \right) 
({r_{+,q}})^{16}{N_{q}}^{3}-36\left(6(1+3\pi {Q}^{2}){N_{q}}+\pi \right) 
({r_{+,q}})^{14}{N_{q}}^{2}\nonumber\\
&+&\left( 108 \left( 7N_{q}+\pi  \right) {N_{q}}^{2}{Q}^{2}-90{N_{q}}
^{2}-12\pi N_{q} \right) ({r_{+,q}})^{12}\nonumber\\
&+& \left( 252\pi {N_{q}}^{3}{Q}^{4}+72 \left(13{N_{q}}+{\frac{5\pi}{3}} 
\right) {Q}^{2}+24N_{q}+2\pi  \right) ({r_{+,q}})^{10}\nonumber\\
&-& \left(12\left(180{N_{q}}+29\pi \right)N_{q} {Q}^{4}+24\left(10N_{q}+
{\frac{5\pi}{6}}\right){Q}^{2} \right)({r_{+,q}})^{8}\nonumber\\
&+& 12\left( 18\pi N_{q}{Q}^{6}+ \left( 52N_{q}+5\pi  \right) {Q}^{4} \right) 
({r_{+,q}})^{6}
- 6\left( 10\left(3N_{q}+\pi  \right) {Q}^{2}+1 \right) ({r_{+,q}})^{4}{Q}
^{4}\nonumber\\
&+& 6{Q}^{6}\left( 3\pi {Q}^{2}-4 \right) ({r_{+,q}})^{2}+18{Q}^{8},\\ 
B_{q}&=&9({r_{+,q}})^{8}{N_{q}}^{2}+6N_{q}({r_{+,q}})^{6}-(1+30N_{q}{Q}^{2})
({r_{+,q}})^{4}+6{Q}^{2}({r_{+,q}})^{2}-3{Q}^{4}.\end{eqnarray}
 \begin{figure}[h!]
 \begin{center}$
 \begin{array}{cccc}
\includegraphics[width=50 mm]{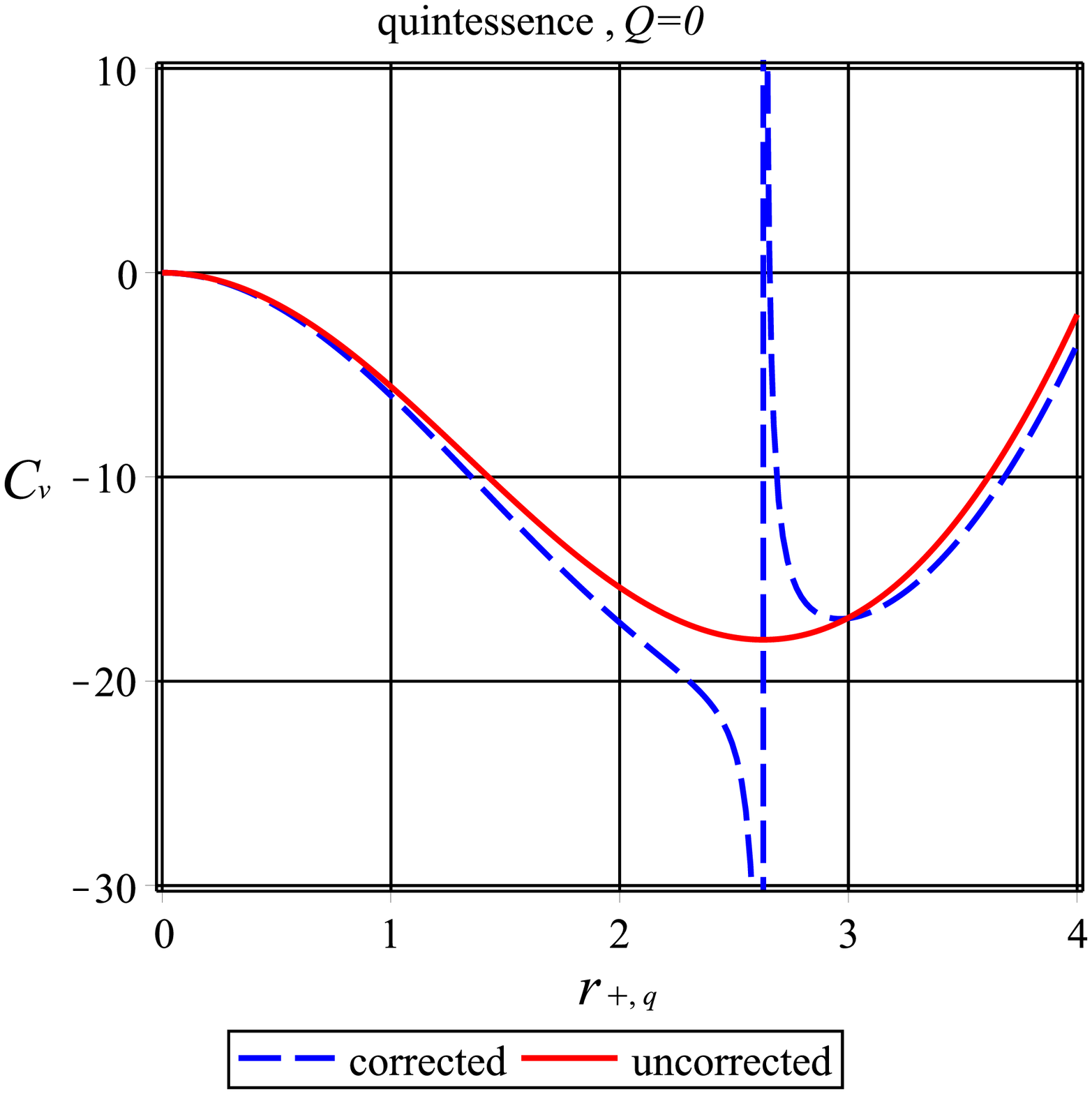}&\includegraphics[width=50 mm]{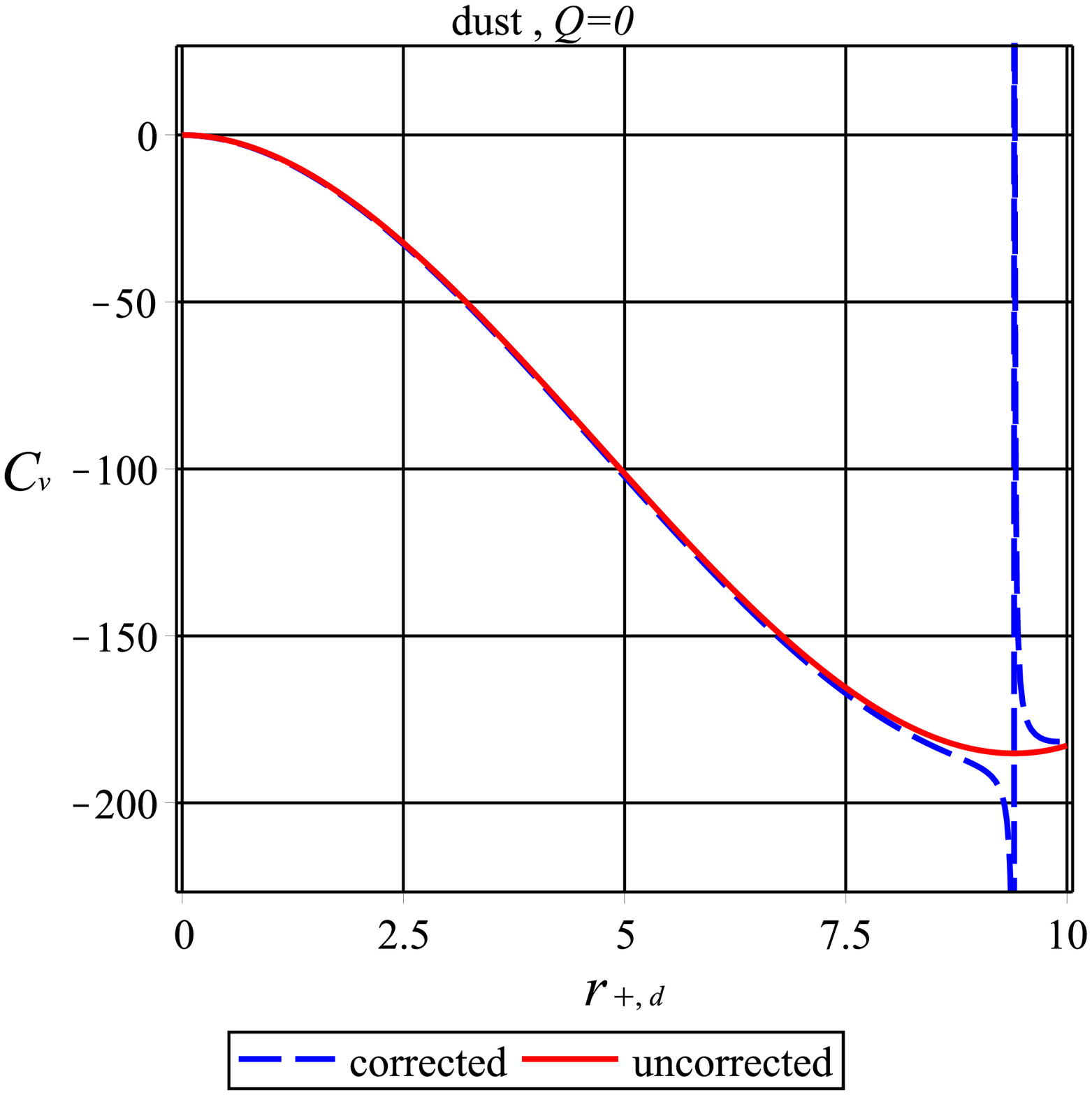}
&\includegraphics[width=50 mm]{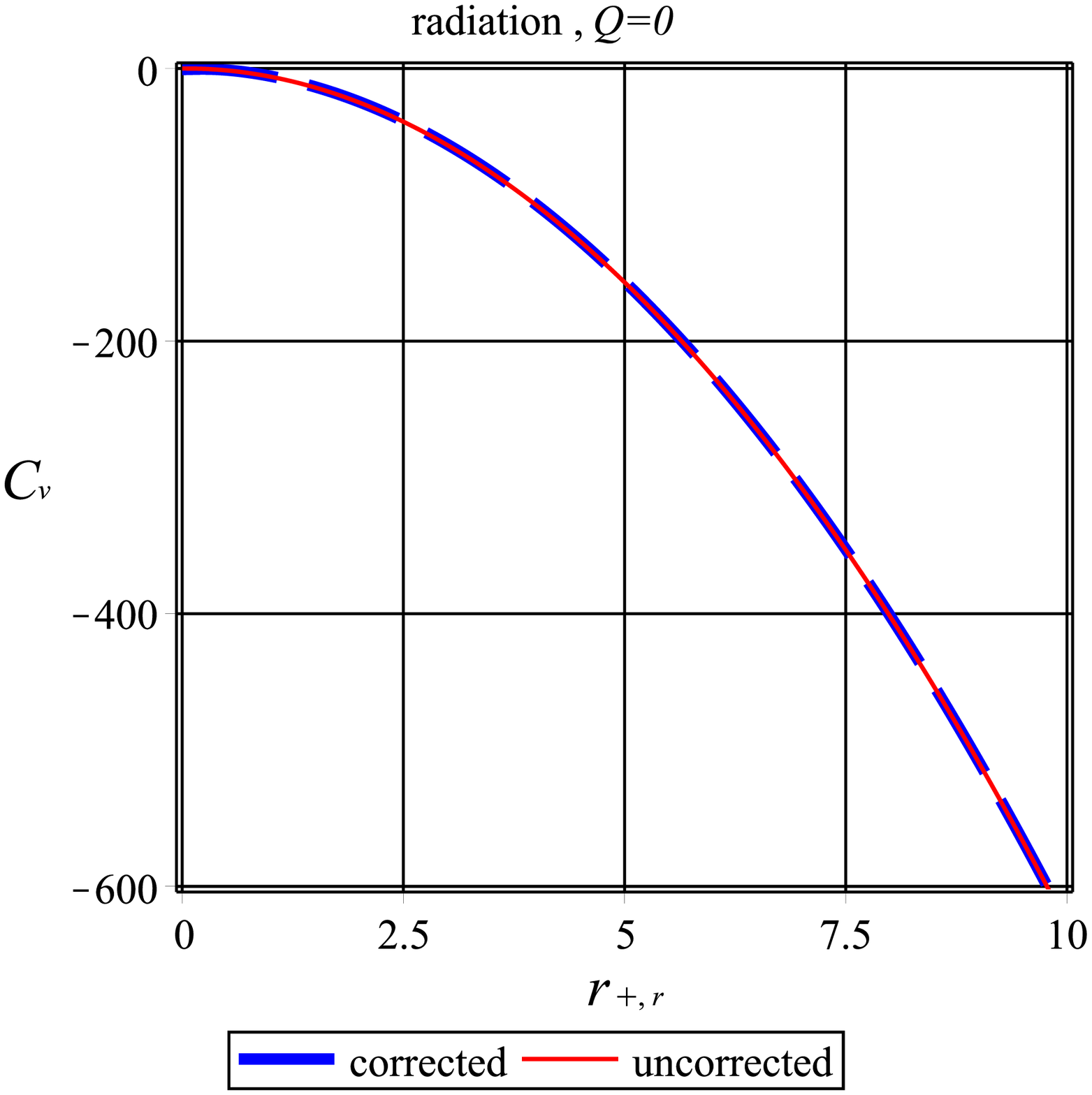}\\
\includegraphics[width=50 mm]{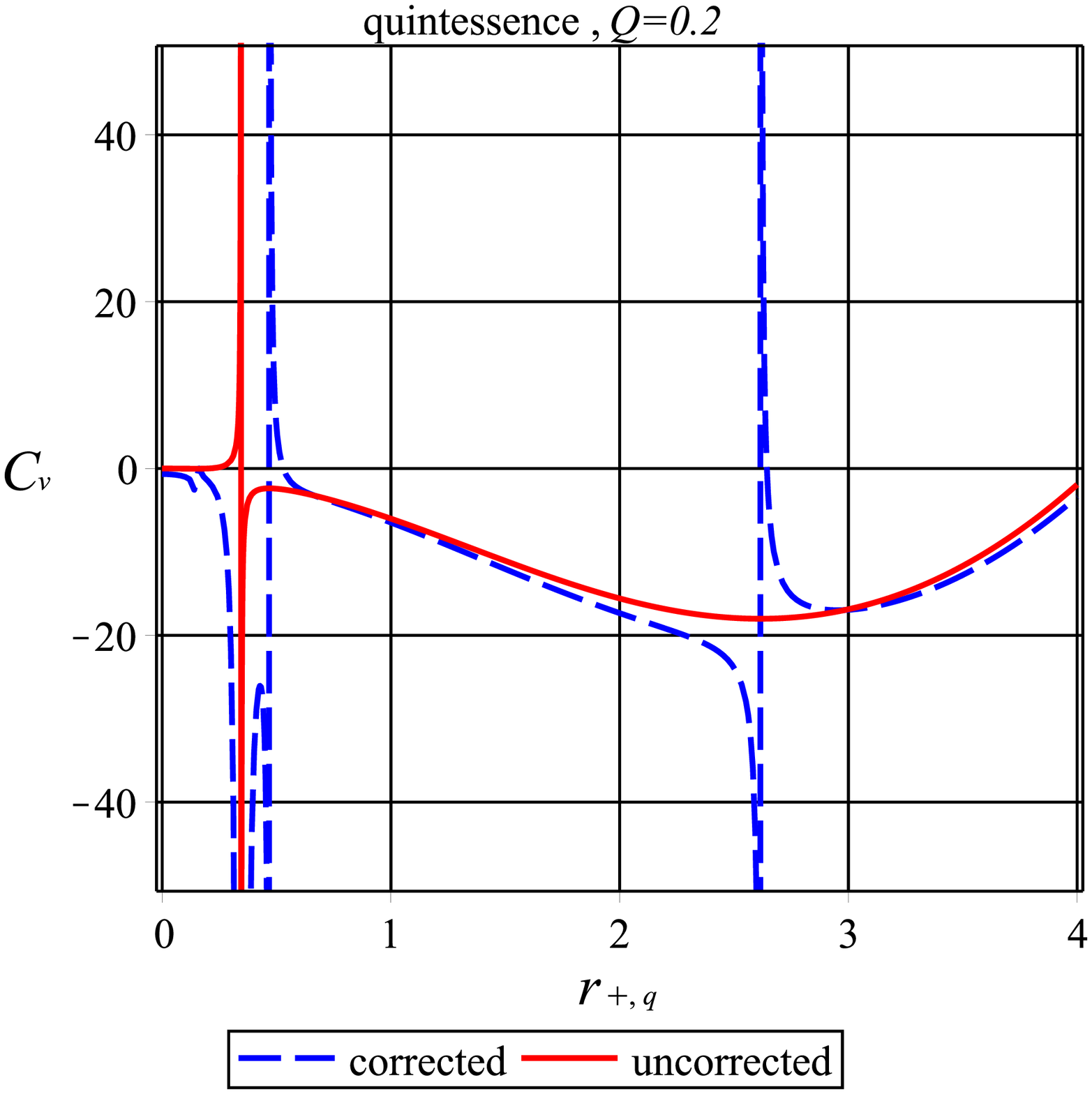}&\includegraphics[width=50 mm]
{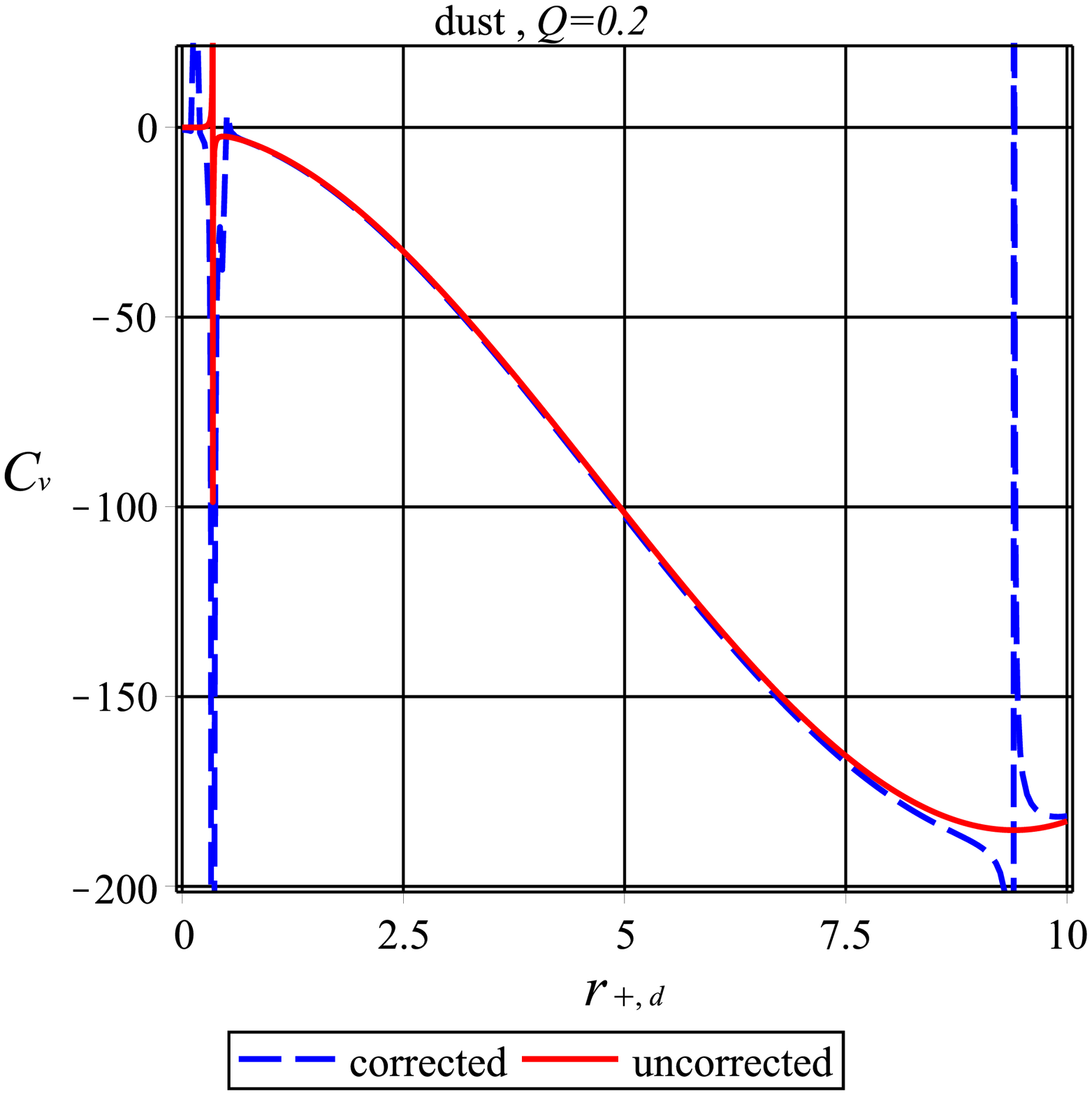}&\includegraphics[width=50 mm]{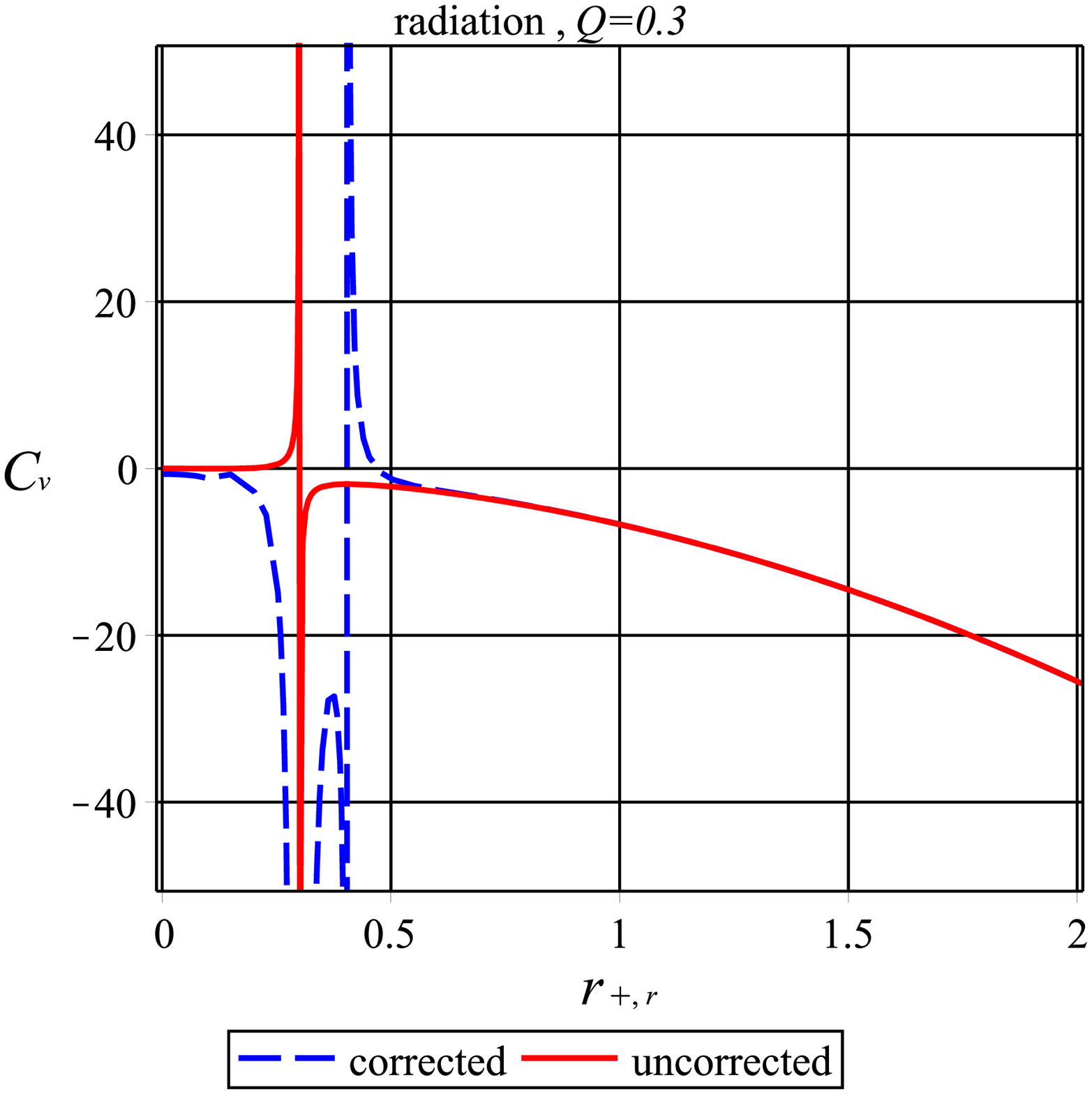}
 \end{array}$
 \end{center}
\caption{Typical behaviour of the specific heat in terms of horizon radius, 
for $N_{s}=0.02$.}
 \label{fig6}
\end{figure}
In the plots of the Fig. \ref{fig6}, we draw specific heat in terms of the 
black hole horizon radius to see effects of thermal fluctuations. We perform 
a graphical analysis for both charged and uncharged cases. In the case of 
uncharged black hole surrounded by quintessence field we can see a completely 
unstable black hole without any phase transition in the valid regions (see 
left 
plots of the Fig. \ref{fig6}). Also, in this case, due to the thermal 
fluctuation 
 the phase transition occurs. Similar result is obtained for the case of 
charged black hole surrounded by quintessence field in Rastall theory, 
however, this time the thermal fluctuation causes the second 
phase transition. So there is the first  order phase transition in both 
corrected and uncorrected thermodynamics.

\begin{figure}[h!]
 \begin{center}$
 \begin{array}{cccc}
\includegraphics[width=50 mm]{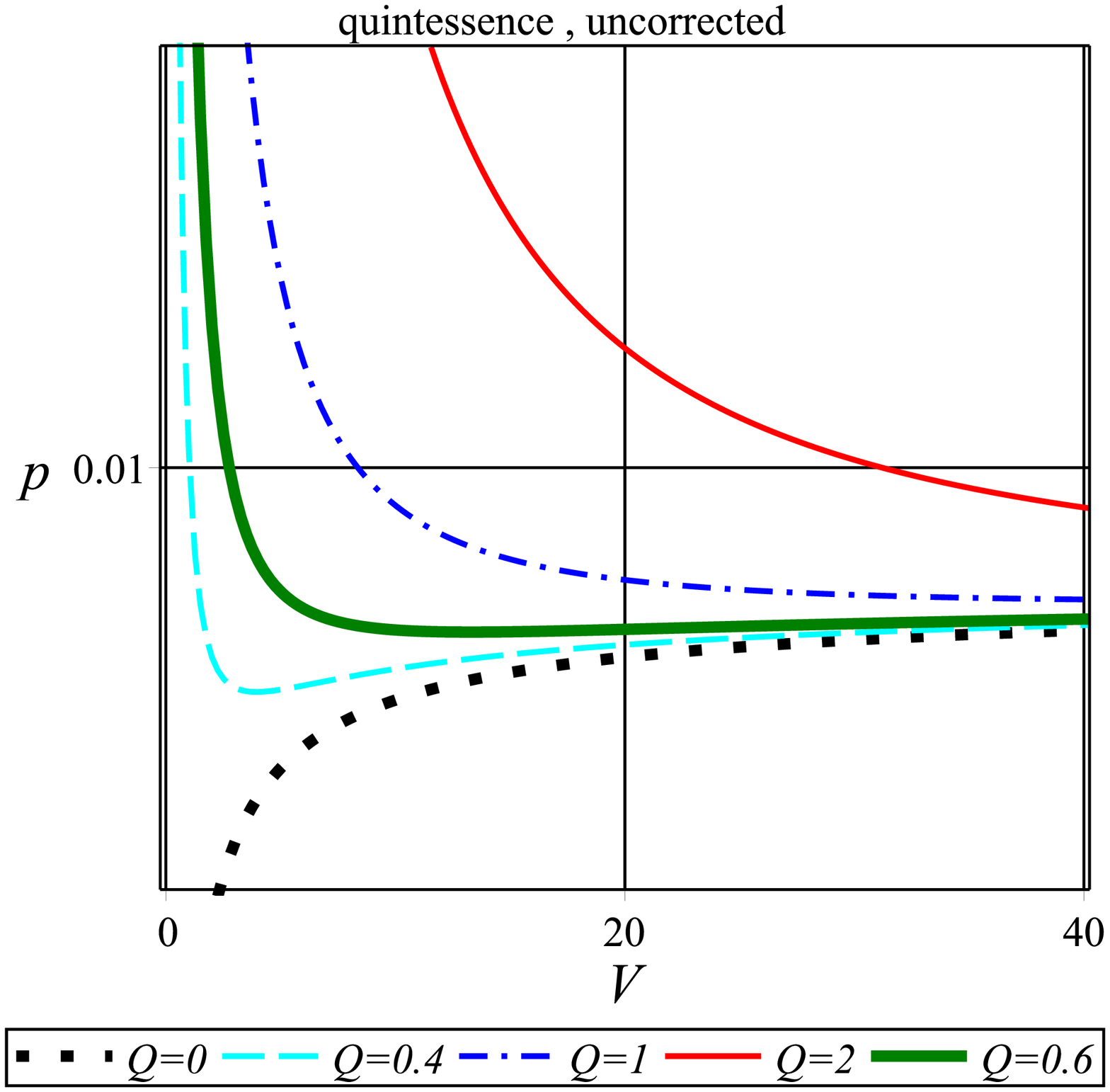}&\includegraphics[width=50 mm]{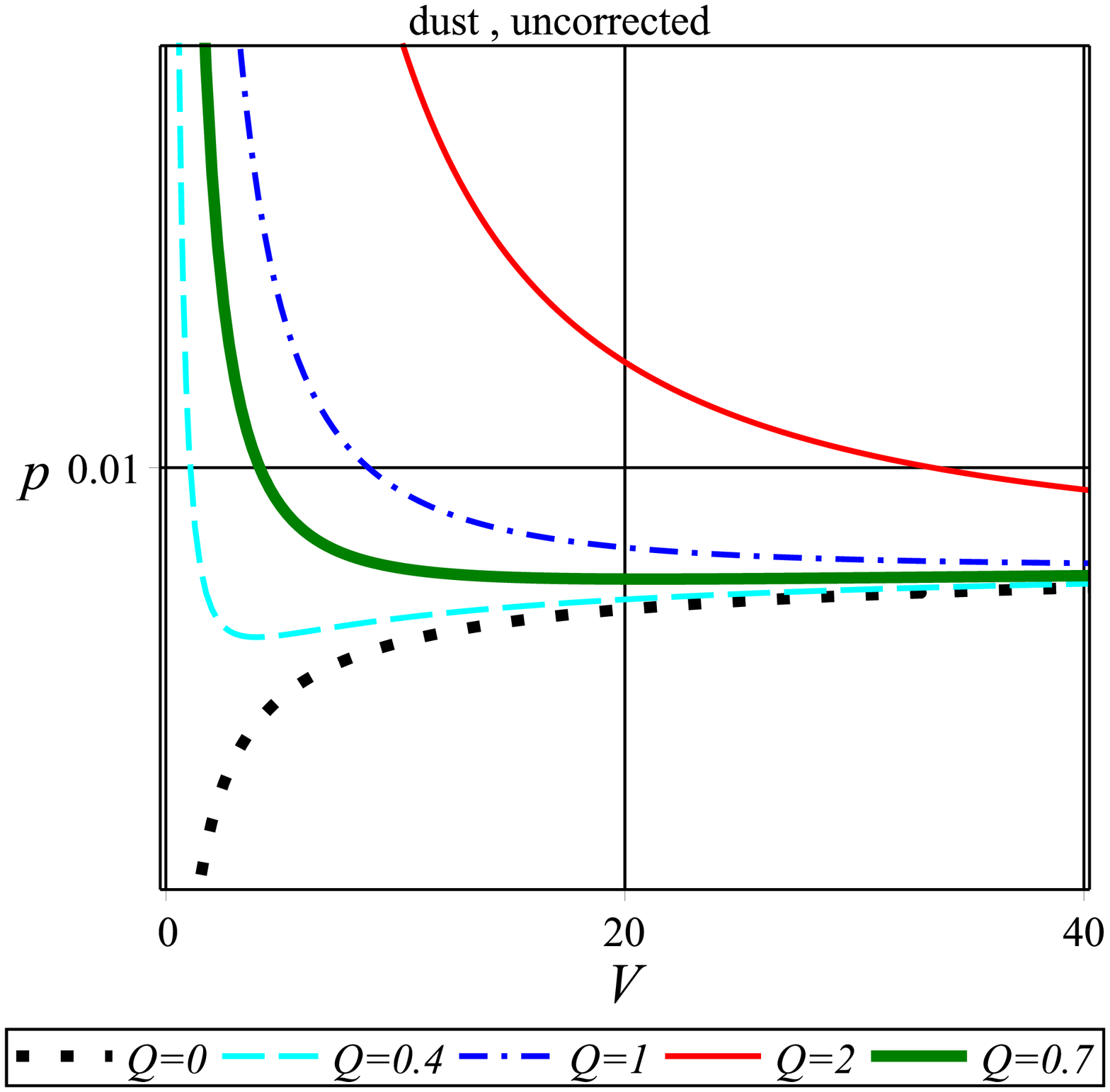}&\includegraphics[width=50 mm]{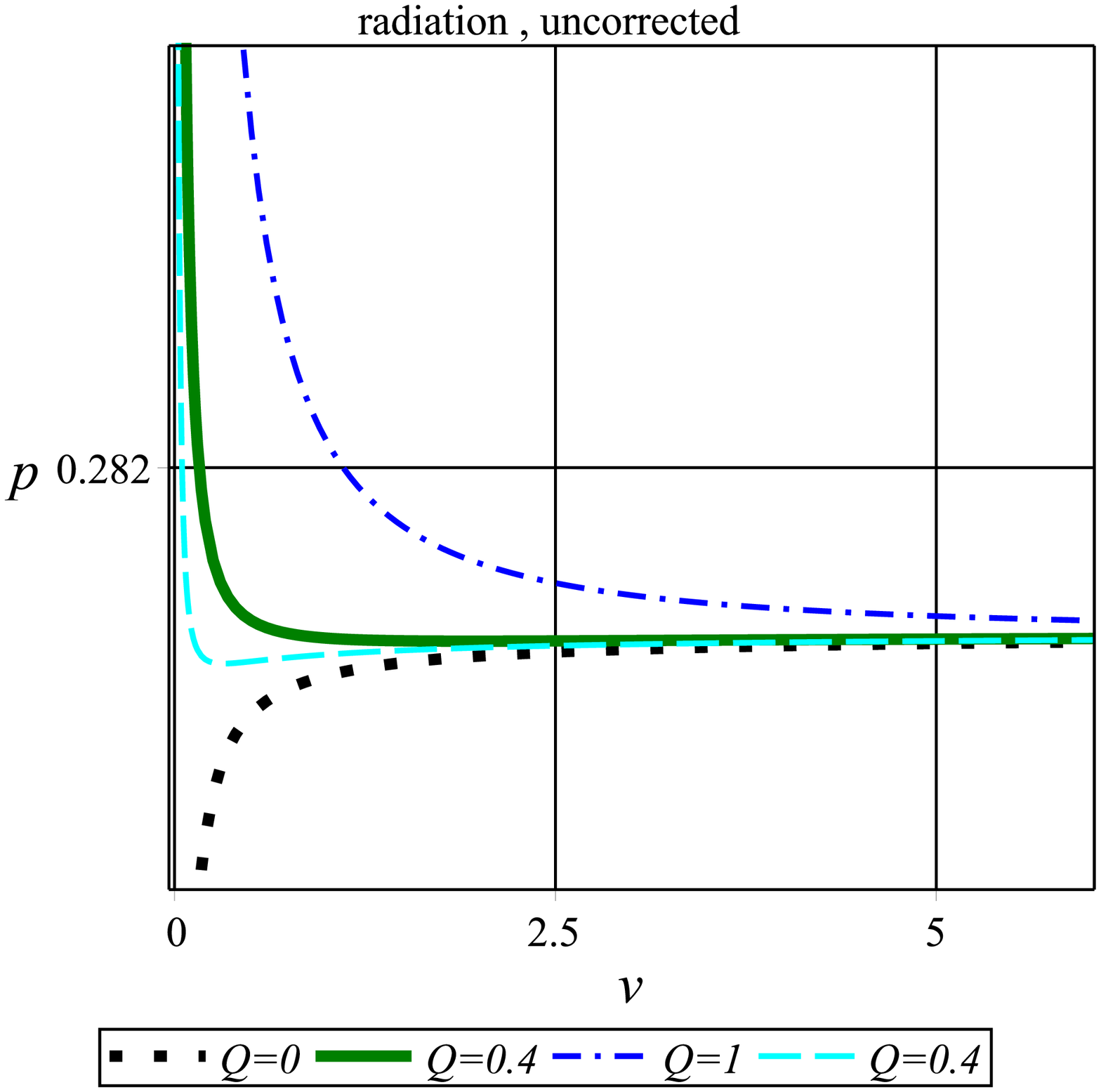}\\
\includegraphics[width=50 mm]{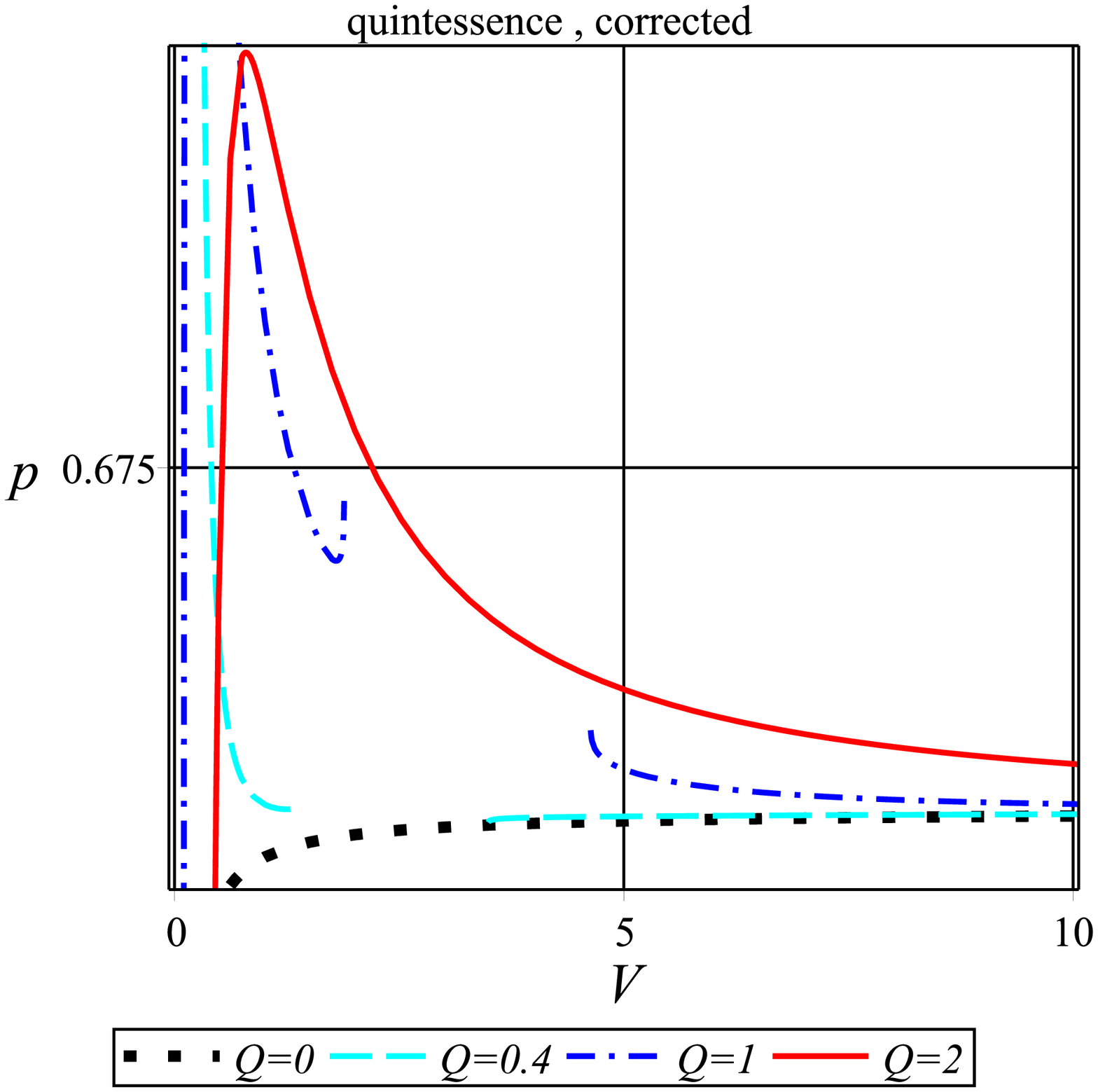}&\includegraphics[width=50 mm]{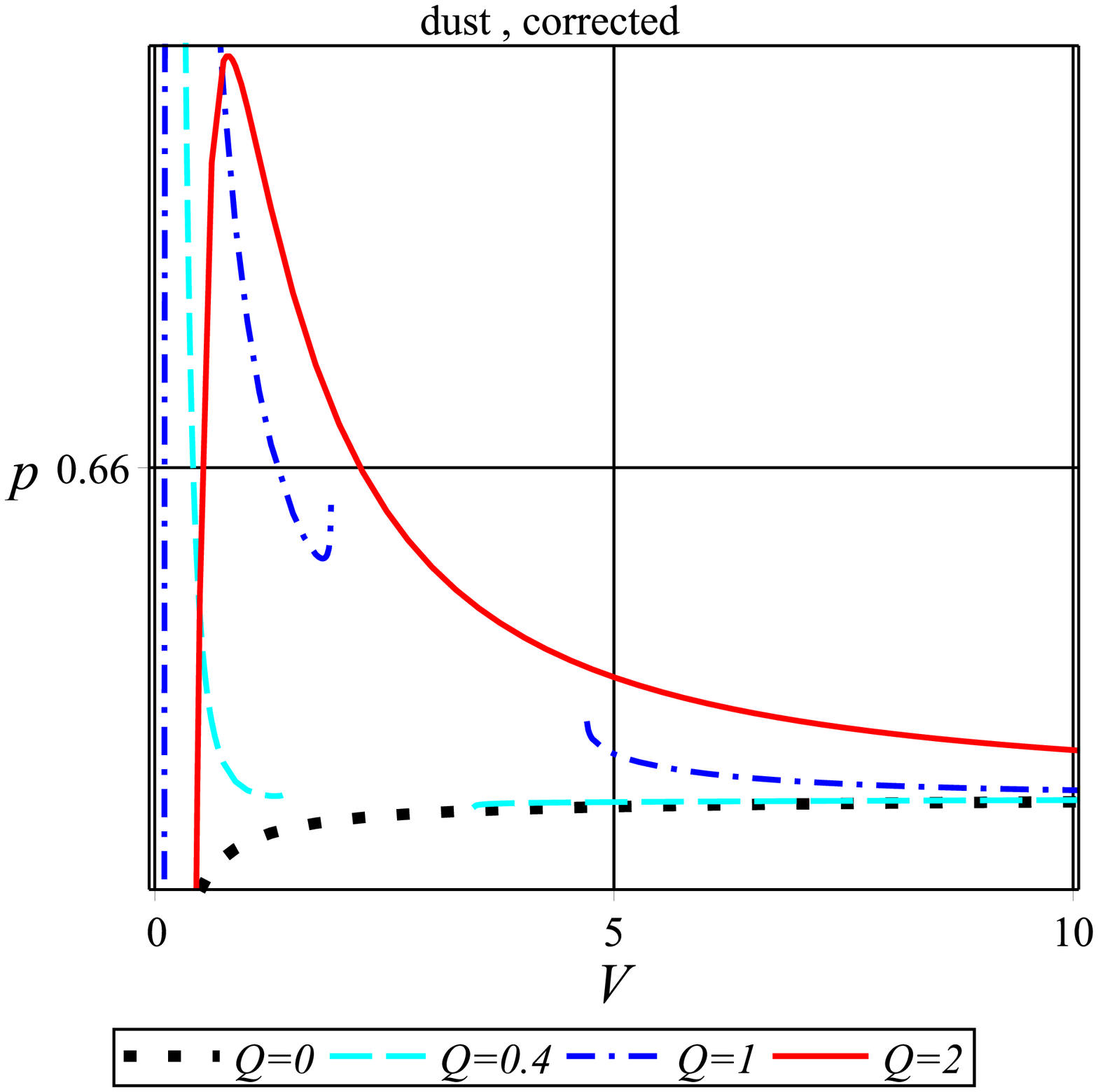}&\includegraphics[width=50 mm]{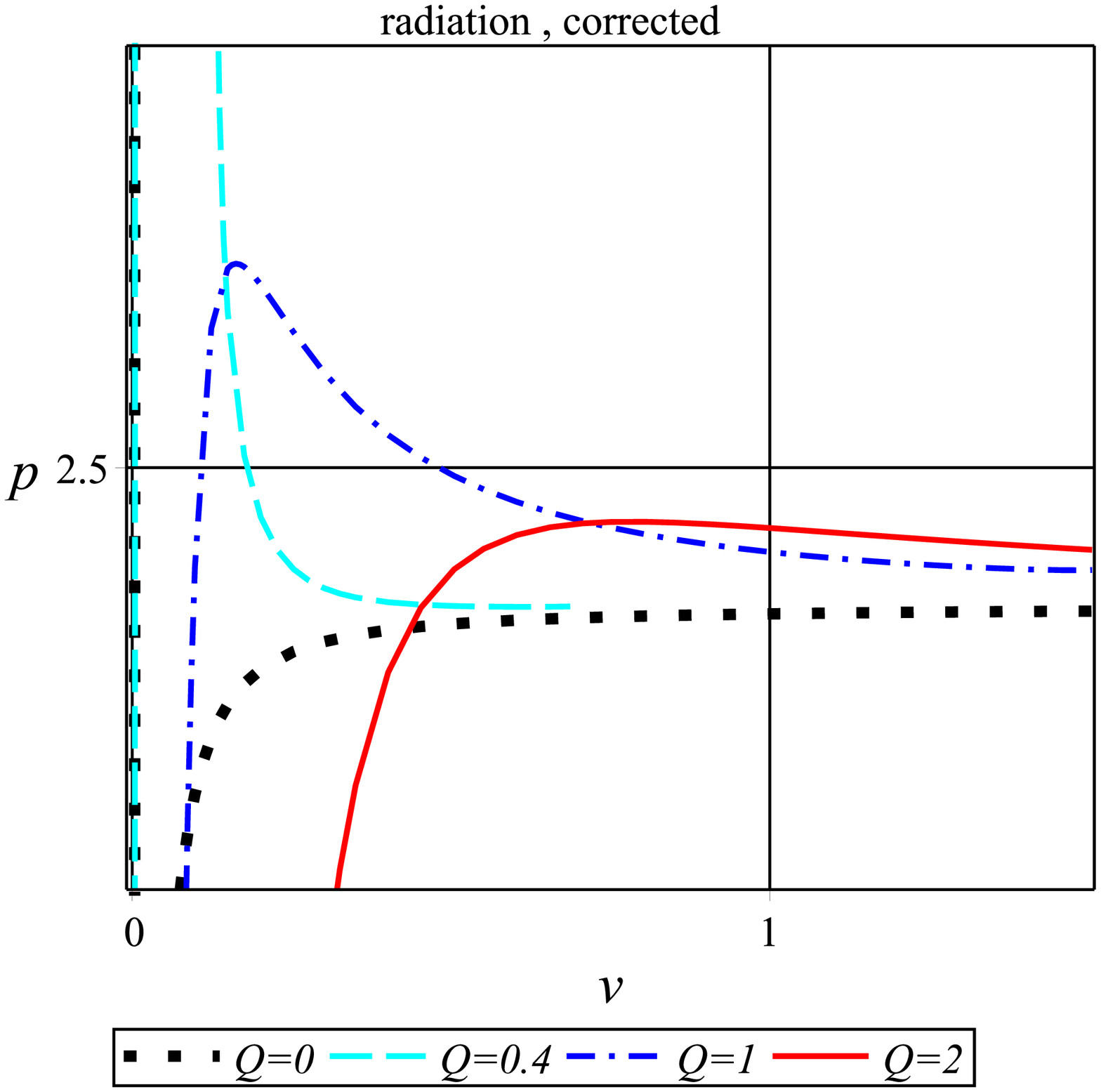}
 \end{array}$
 \end{center}
\caption{$p-V$ diagram, for $N_{s}=0.02$.}
 \label{fig7}
\end{figure}

Our numerical analysis about critical points is presented in Fig. \ref{fig7}. We 
plot $p$ in terms of $V$ for both cases of corrected (lower plots) and 
uncorrected (upper plots) black hole entropy. In the case of black hole 
surrounded by quintessence field in Rastall theory, we find that points 
get affected by thermal fluctuations. Neglecting thermal fluctuations (upper left 
plot of the Fig. \ref{fig7}), we have critical points which are denoted by thick 
solid green line of the upper left plot of the Fig. \ref{fig7}. Here, 
we can see Van der Waals like behaviour. However, in presence of the thermal 
fluctuations, there is no critical point or Van der Waal behaviour. Instead 
we can see a maximum for the pressure of highly charged black hole.

\subsection{Dust field}
Now, we consider black hole surrounded by dust field in Rastall theory. In absence of thermal fluctuations we can obtain the following specific heat:
\begin{equation}
C_{V}=\frac {2({r_{+,d}})^{2} \left( \pi ({r_{+,d}})^{2}-2N_{d}\sqrt{\pi}({r_{+,d}})^{3}-\pi{Q}^{2} \right) }{3{Q}^{2}-({r_{+,d}})^{2}}.
\end{equation}
In presence of thermal fluctuations, we have a complicated expression like (\ref{CQ}), hence, we only give graphical analysis. Similar to previous  case we can see that due to the thermal fluctuation  an existence of the second order phase transition (see middle plots of the Fig. \ref{fig6}). Upper plots of the Fig. \ref{fig6} are corresponding to the uncharged cases, while lower plots show charged black holes.

Now, by using the equation (\ref{p}), one can calculate pressure in terms of event horizon radius. Then, by using the relation (\ref{V}), we represent pressure in terms of volume. In absence of thermal fluctuations, we get 
\begin{equation}
pV^{3}+f(V)=0.
\end{equation}
If $f(V)=a_{1}V^{2}+a_{2}V+a_{3}$, then we have Van der Waal  equation of state, where $a_{1}$, $a_{2}$ and $a_{3}$ are some constants. However, in this case, we find that 
\begin{equation}
f(V)\approx0.05 V^{\frac{7}{3}}-0.4 Q^{2}V^{\frac{5}{3}},
\end{equation}
which behaves approximately like Van der Waals equation of state. It is interesting to see that this is independent of $N_{d}$. This is illustrated by top middle plot of the Fig. \ref{fig7}. Here, both conditions (\ref{critical}) yield  to the following equation:
\begin{equation}
Y_{d}\equiv V_{c}^{\frac{5}{3}}-\frac{4}{3}V_{c}^{\frac{2}{3}}+ \left({\frac {10}{21}}  -\frac{5}{7}V_{c}\right) {Q}^{2}{\pi }^{\frac{2}{3}}(48)^{\frac{1}{3}}=0,
\end{equation}
where $V_{c}$ is critical volume which yields to the critical horizon radius $r_{+,d,c}$. 

\begin{figure}[h!]
 \begin{center}$
 \begin{array}{cccc}
\includegraphics[width=70 mm]{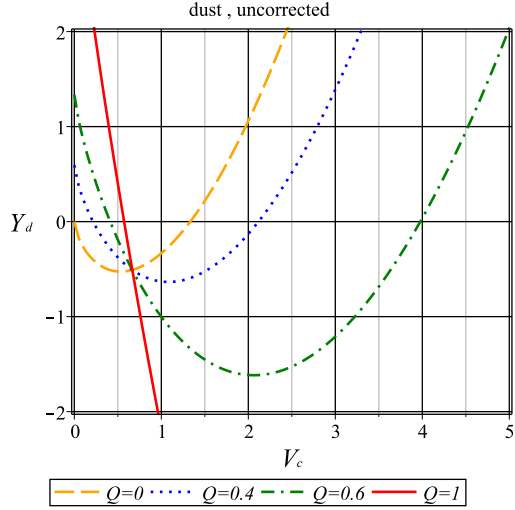}
 \end{array}$
 \end{center}
\caption{Some possible values of critical volume of black hole surrounded by dust field in Rastall theory in absence of thermal fluctuations.}
 \label{fig8}
\end{figure}

In the Fig. \ref{fig8}, we have possible values of critical volume of black hole surrounded by dust field in Rastall theory in absence of thermal fluctuations. For example, by choosing $Q=0.6$ one can obtain $V_{c}=4$ which means $r_{+,d,c}\approx1$. It may be larger or smaller than the event horizon radius which  also depends on $M$ and $N_{d}$.\\
In presence of thermal fluctuations, we can see similar behaviour with the previous case (quintessence field), where there is no critical point or Van der Waal  like behaviour.

\subsection{Radiation field}
Finally, we discuss about stability of the black hole surrounded by radiation field in Rastall theory.
In absence of thermal fluctuations, the specific heat has following form:
\begin{equation}
C_{V}={\frac {2\pi ({r_{+,r}})^{2} \left(({r_{+,r}})^{2}+3N_{r} -{Q}^{2}\right) }{{Q}^{2
}-({r_{+,r}})^{2}-9N}},
\end{equation}
while in presence of the thermal fluctuation, specific heat modified non-trivially if black hole charge is large. It is clear that the specific heat is zero if $({r_{+,r}})^{2}+3N_{r} -{Q}^{2}=0$. From the top right plot of the Fig. \ref{fig6}, we can see that corrected and uncorrected specific heat are similar for the uncharged black hole. 

This is illustrated that uncharged black hole surrounded by radiation field in Rastall theory is completely unstable without any phase transition. Also, we find that for the small charge, the same results are obtained. For the large enough charge, we can see that thermal fluctuation is responsible for instability of black hole.

Finally, we can investigate critical points of this model.
Neglecting thermal fluctuations, we get
\begin{equation}
pV^{3}+g(V)=0,
\end{equation}
where
\begin{equation}
g(V)\approx0.075 V^{\frac{7}{3}}+1.2 (N_{r}-0.3Q^{2})V^{\frac{5}{3}}.
\end{equation}
Here, we note that the critical point exists about $Q=0.4$ (for $N_{r}=0.02$). It is illustrated by green thick line of top middle plot of the Fig. \ref{fig7}. In presence of thermal fluctuations, however, critical points may exist  only for the highly charged black hole.

\section{Geometrothermodynamics}
In this section, following the works of \cite{G1, G2, G3, G4} we study geometric formalism of the thermal system, and investigate the thermodynamics of a charged black hole surrounded by perfect fluids. In that case we could calculate scalar curvature to obtain singular points. Then, we compare it with zeros of specific heat to obtain some information.
\subsection{Quintessence field}
\begin{figure}[h!]
 \begin{center}$
 \begin{array}{cccc}
\includegraphics[width=50 mm]{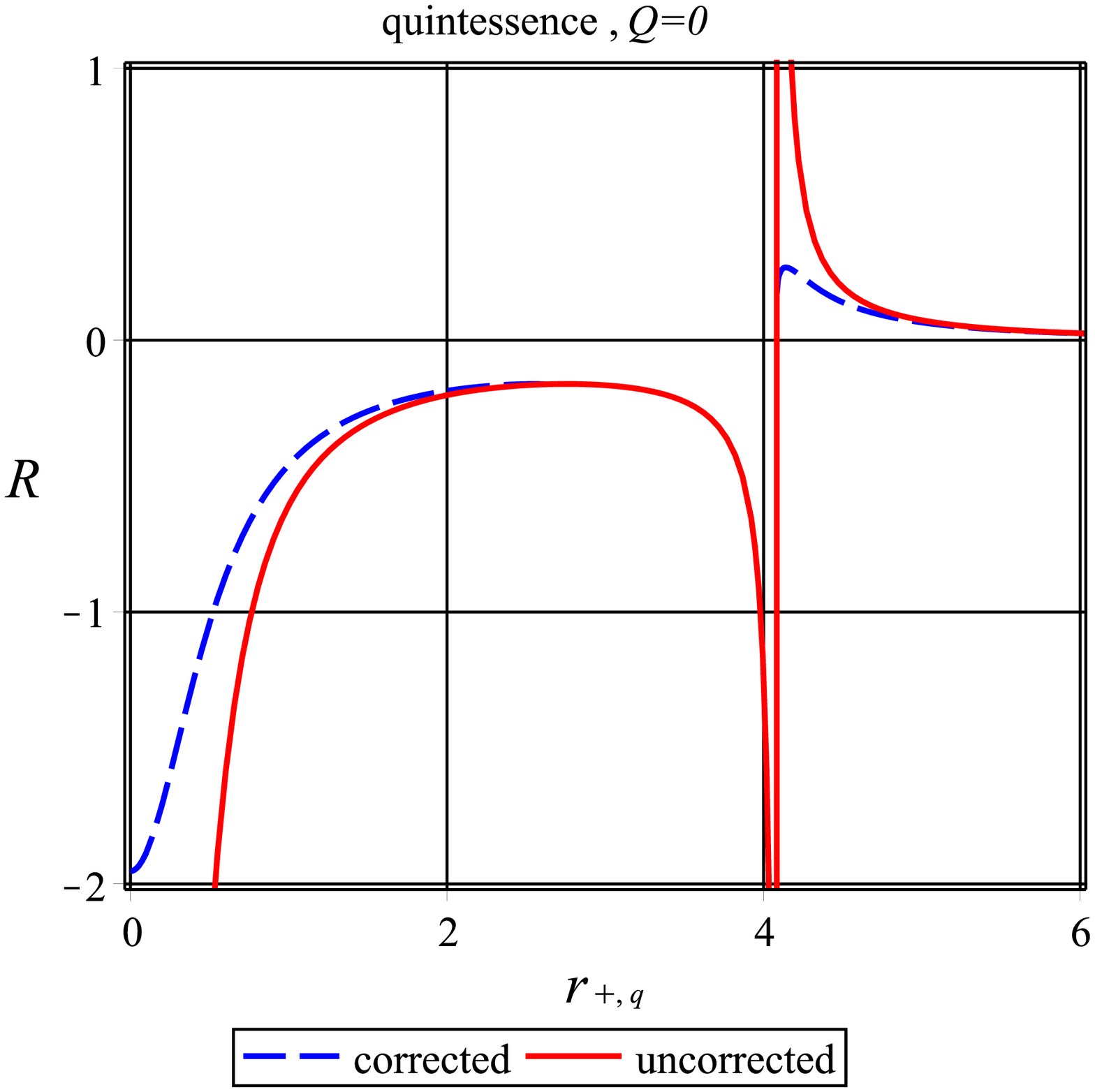}&\includegraphics[width=50 mm]{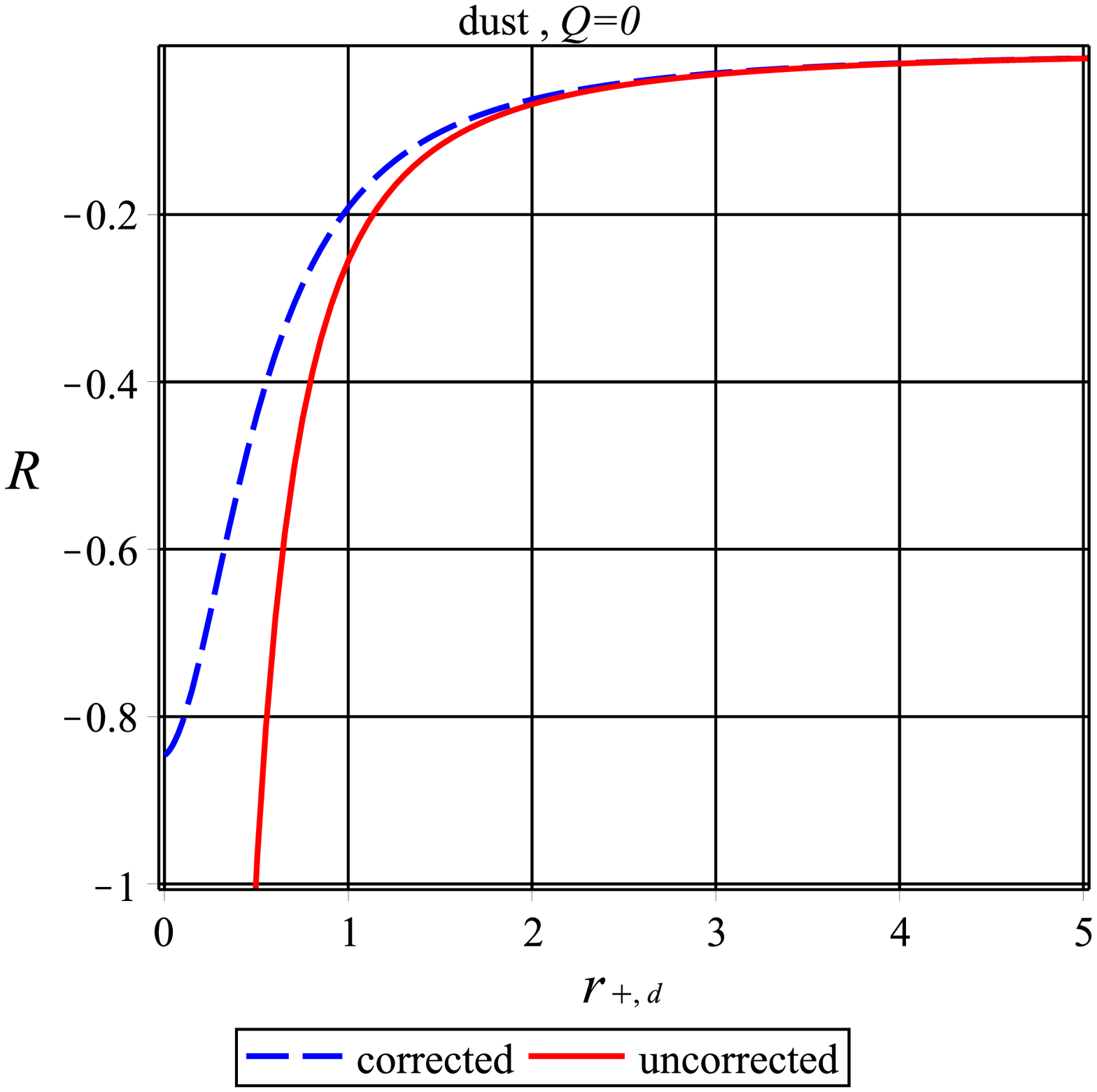}&\includegraphics[width=50 mm]{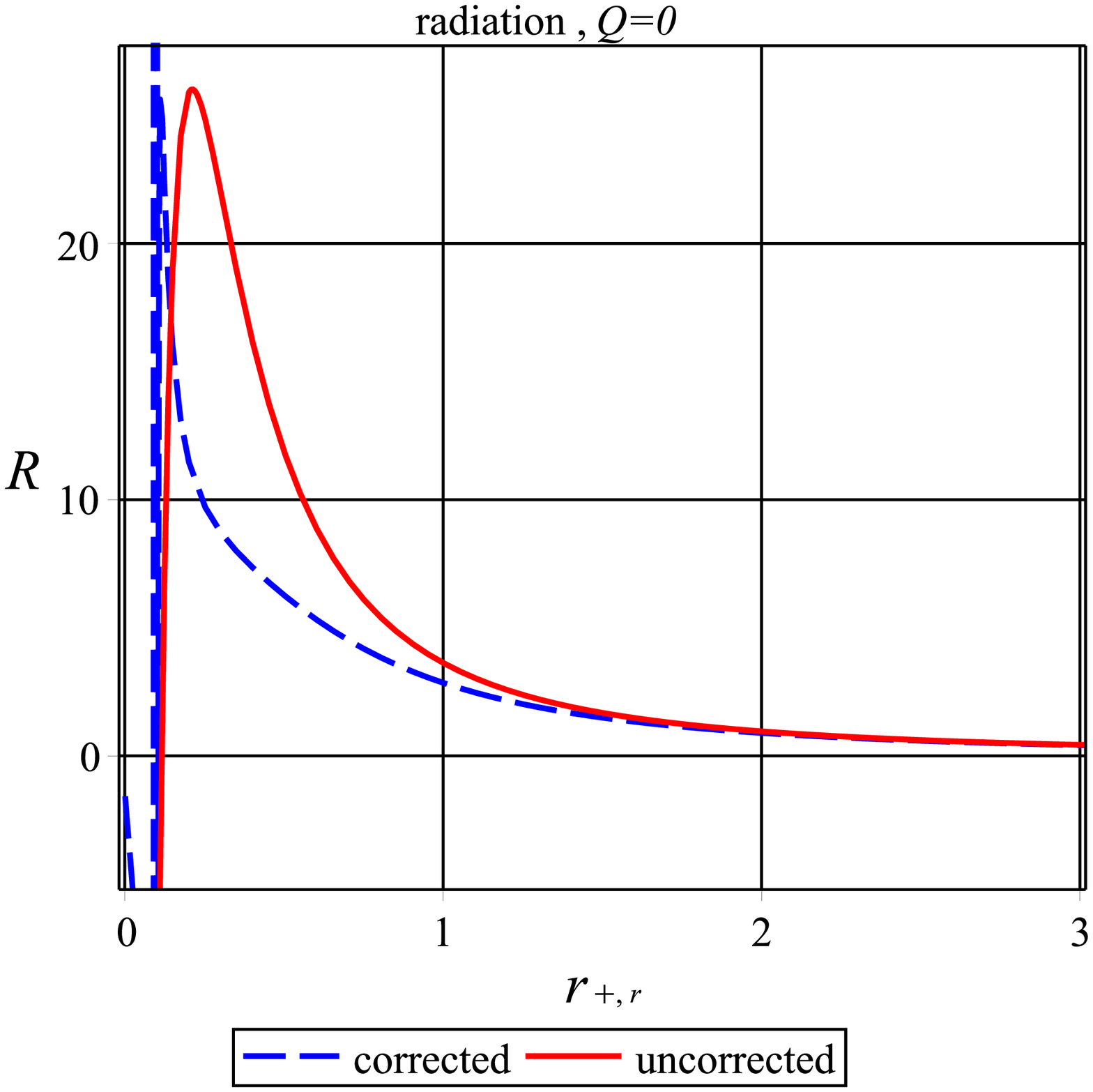}\\
\includegraphics[width=50 mm]{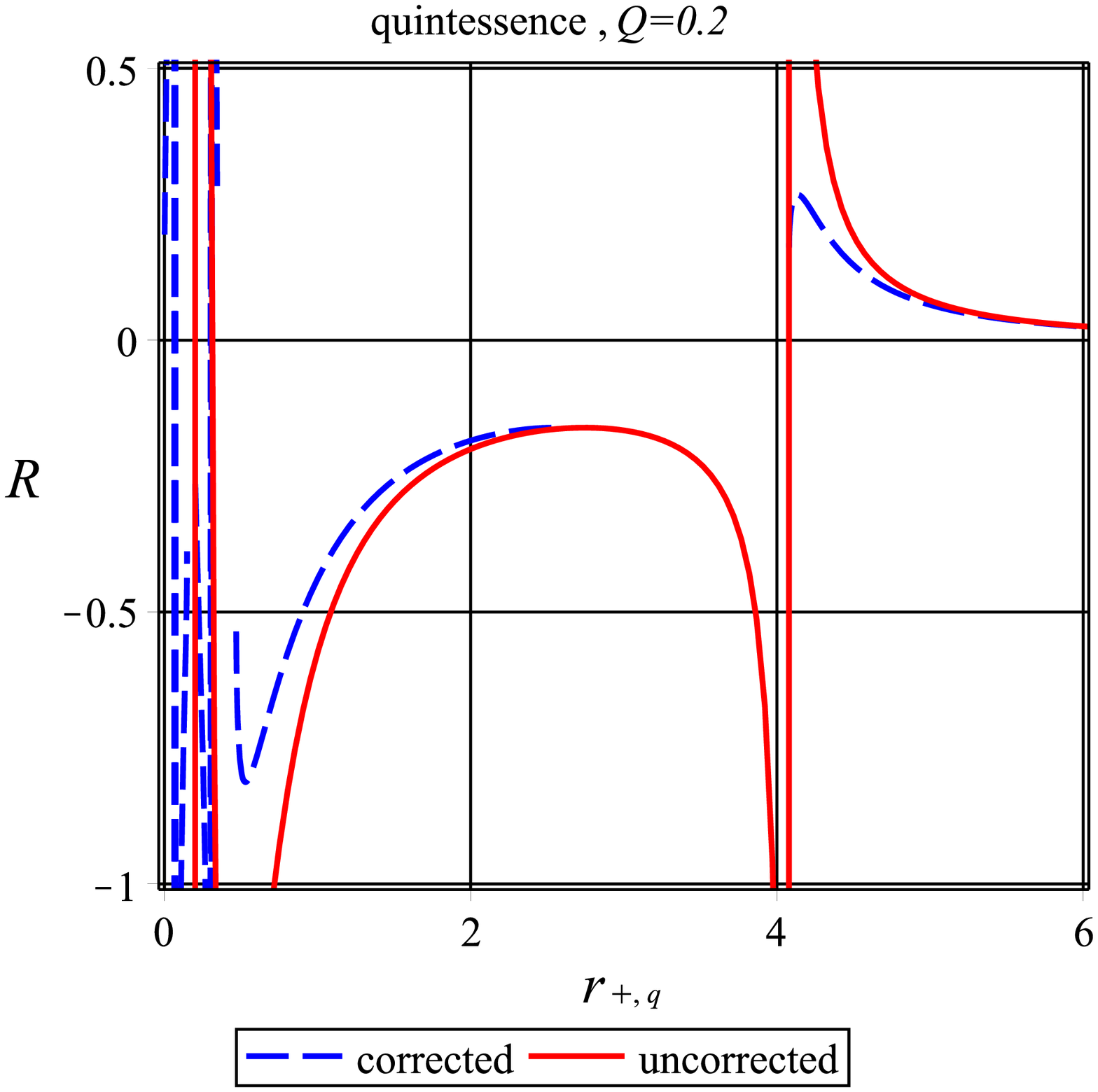}&\includegraphics[width=50 mm]{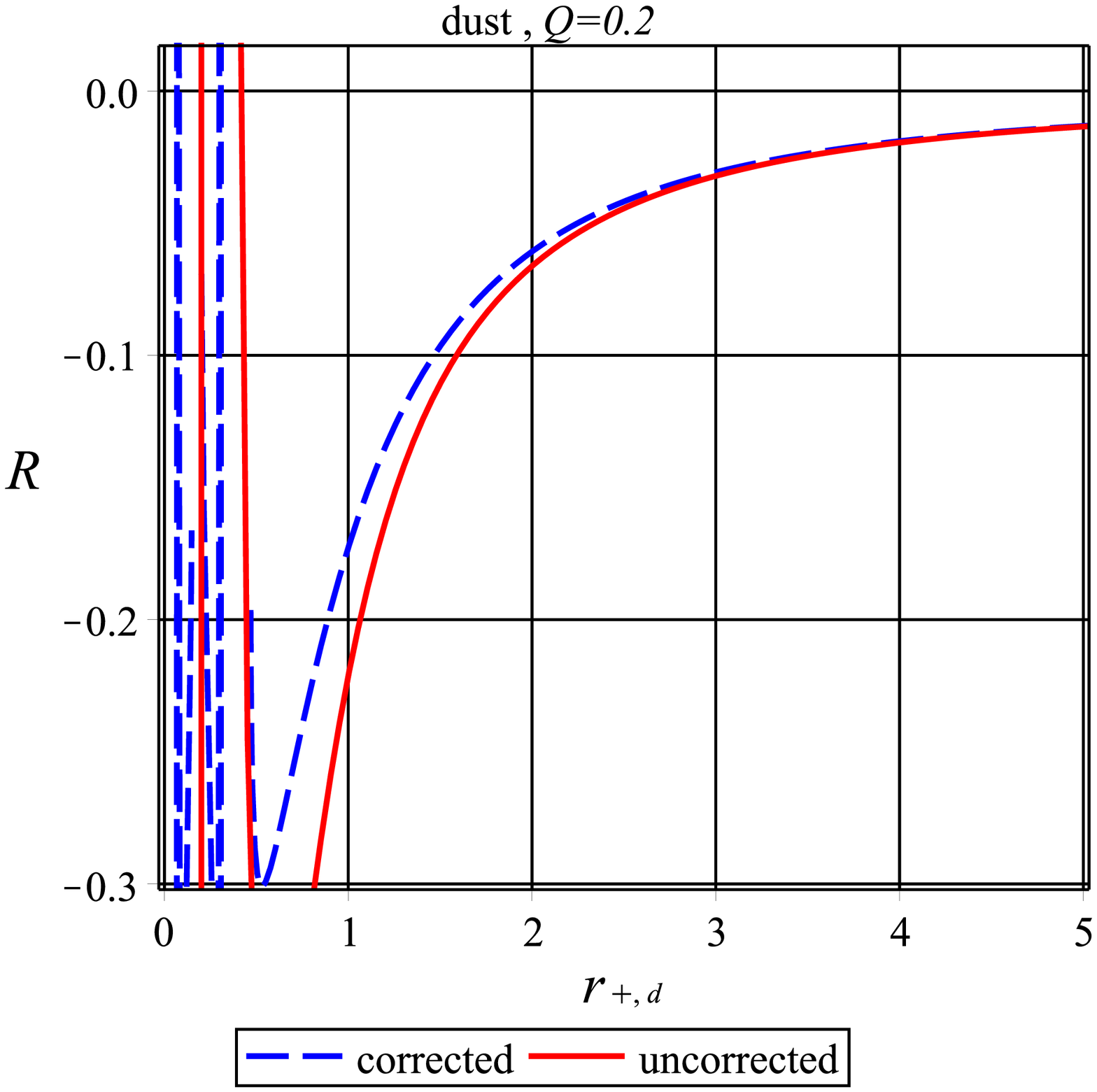}&\includegraphics[width=50 mm]{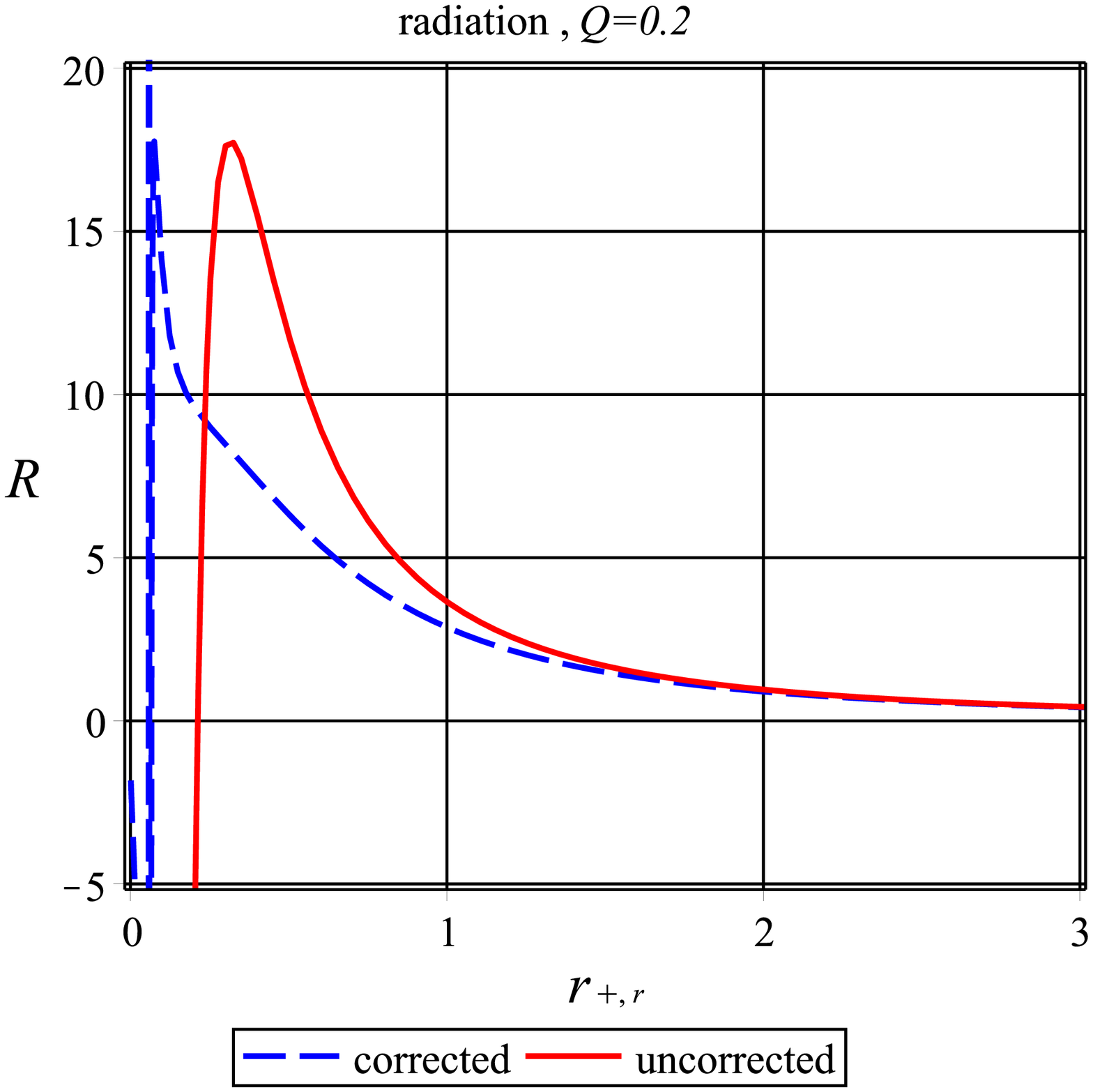}
 \end{array}$
 \end{center}
\caption{Curvature scalar of Ruppeiner metric corresponding to a black hole surrounded by perfect fluid in terms of horizon radius for $N_{s}=0.02$.}
 \label{fig9}
\end{figure}

In the Ref. \cite{sud0}, the relevant scalar curvature of Ruppiner formalism is found as 
\begin{equation}
R=-\frac{17\pi^{2}Q^{2}-9{\mathcal{S}}^{2}N_{q}-7\pi \mathcal{S}}{4S(\pi^{2}Q^{2}+3{\mathcal{S}}^{2}N_{q}-\pi \mathcal{S})},
\end{equation}
where $\mathcal{S}$ is given by the equation (\ref{mainq}). This equation still holds as the black hole temperature is  not affected by thermal fluctuations and the first law of thermodynamics is valid in presence of corrections also. From the left plots of the Fig. \ref{fig9}, it is clear that there is at least one singular point which coincides with zero of specific heat represented by left plots of the Fig. \ref{fig6}. In the case of charged black hole surrounded by quintessence field, we can obtain information about the first and the second order phase transitions from asymptotic behaviours of scalar curvature. But in the case of uncharged black hole we have no information about the first order
phase transition.

\subsection{Dust field}
In the case of black hole surrounded by dust field, we have
\begin{equation}
R=-\frac{13\pi^{2}Q^{2}-4\sqrt{\pi}{\mathcal{S}}^{\frac{3}{2}}N_{d}-3\pi \mathcal{S}}{4{\mathcal{S}}(\pi^{2}Q^{2}+2\sqrt{\pi}{\mathcal{S}}^{\frac{3}{2}}
N_{q}-\pi \mathcal{S})},
\end{equation}
where  $\mathcal{S}$ is same corrected entropy as given in (\ref{maind}). Its behaviour is illustrated by middle plots of the Fig. \ref{fig9} which 
confirm that there is no singular point if $Q=0$. However, charged black hole has singular point which coincides with phase transition point of the black hole (see Fig. \ref{fig6}).
\subsection{Radiation field}
The curvature scalar for the case of radiation field is given by
\begin{equation}
R=\frac{39\pi Q^{2}-49{\mathcal{S}}+33\pi N_{r}}{4{\mathcal{S}}(\pi Q^{2}-{\mathcal{S}}-3\pi N_{r})}.
\end{equation}
Here, also $\mathcal{S}$  has same expression as mentioned in (\ref{maind}). The right plots of the Fig. \ref{fig9} show that there is no singular point, otherwise there is a maximum which is near the phase transition point.

\section{Conclusion}
We have considered a charged black hole surrounded by perfect fluid in Rastall theory. Thermodynamics of such system is already been studied in the Ref. \cite{sud0} which assumes quintessence or dust fields as the perfect fluid. On the other hand, we have considered in addition radiation field also as the perfect fluid and studied effects of thermal fluctuations for the system. Thermal fluctuations become important when the size of black hole  decreases due to the Hawking radiation and, hence, this is quite anticipated that   black hole stability and critical points must also be affected. Therefore, in each section, we performed our analysis separately for quintessence, dust and radiation fields. First of all, we have reviewed some important properties (specially thermodynamics properties) of the system and obtained appropriate condition to verify the first law of thermodynamics. 
Here, we have found that, in the case of constant $Q$ and $N_{s}$, the first law of thermodynamics holds in absence of thermal fluctuations. However, this condition gets modified in presence of thermal fluctuation. We have also discussed about the horizon structure of the model and found outer horizon which denoted by $r_{+}$.

 Moreover, we have obtained the corrected entropy. Correction term, which is logarithmic in nature, comes from  the thermal fluctuations. We found that there is no differences between entropy of the charged black hole surrounded by quintessence, dust or radiation fields when the horizon radius is small. On the other hand, for the case of uncharged black hole, the entropy of black hole surrounded by radiation field has different behaviour at the small radius. With the help of corrected entropy,  we have explored modified thermodynamics of the system as described above. We have calculated Helmholtz free energy and specific heat in the separate sections. Because some expressions are too large in presence of logarithmic correction, we give graphical analysis to see thermal fluctuation effects. Although the entropy of three different cases are approximately similar, but the behaviours of the Helmholtz free energy are quite different. In the case of uncharged black hole surrounded by quintessence, dust and radiation fields, the effect of thermal fluctuations may be increasing, decreasing or both (depending upon values of the event horizon radius) for the Helmholtz free energy. In the case of charged black hole surrounded by quintessence field, the   thermal fluctuations   increases free energy with the same general behaviour (linear variation with radius). On the other hand, in the cases of dust and radiation field the effect of logarithmic correction is decreasing for the Helmholtz free energy. In all cases, there exists a minimum for the radius which 
 falls outside the region for which Helmholtz free energy  is not defined. We can say that   the black hole instability at small radius occurs due to the thermal fluctuations. It means that by decreasing black hole size, we have 
 transition from stable to  unstable phase due to thermal fluctuations. As discussed by the Ref. \cite{sud0}, the charged black hole surrounded by perfect fluid has one type of phase transition which obtained by analysing the specific heat. Here,  we have shown that presence of logarithmic term is responsible for the second order phase transition. Hence, region of the stability gets modified due to the thermal fluctuation. By analysing critical points, we have found that the charged black hole surrounded by perfect fluid may behave like Van der Waal  fluid in the absence of thermal fluctuations, i.e., when the event horizon radius  of black hole is large. However, for the small black hole (in presence of thermal fluctuations) there is no Van der Waal like behaviour. We have also created the Ruppiner geometric structure \cite{G1} for a black hole surrounded by quintessence, dust and radiation fields. Here, we have found that only in the case of the charged black hole surrounded by quintessence field, the singular point of scalar curvature   coincides with zero of specific heat. This will also be interesting to
  to consider higher order corrections to the entropy and study the stability condition as discussed in Refs. \cite{higher1, higher2, higher3}.

\section*{Acknowledgments}
B. Pourhassan would like to thanks Iran Science Elites
Federation.


\begin{thebibliography}{}
\bibitem{ras1}P. Rastall, Phys. Rev. D 6, 3357 (1972).
\bibitem{ras2}P. Rastall, Can. J. Phys. 54, 66 (1976).
\bibitem{2} E. R. Bezerra de Mello, J. C. Fabris and B. Hartmann, Class. Quantum Grav. 32, 085009 (2015).
\bibitem{3}A. M. Oliveira, H. E. S. Velten, J. C. Fabris and L. Casarini, Phys. Rev. D 92, 044020 (2015).
\bibitem{4}K. A. Bronnikov, J. C. Fabris, O. F. Piattella and E. C. Santos, Gen. Relativ. Gravit. 48, 162 (2016).
\bibitem{5}A. M. Oliveira, H. E. S. Velten, J. C. Fabris and L. Casarini, Phys. Rev. D 93, 124020 (2016).
\bibitem{6}C. E. M. Batista, M. H. Daouda, J. C. Fabris, O. F. Piattella and D. C. Rodrigues, Phys. Rev. D 85, 084008 (2012).
\bibitem{7}G. F. Silva, O. F. Piattella, J. C. Fabris, L. Casarini and T. O. Barbosa, Grav. Cosmol. 19, 156 (2013).
\bibitem{8}F. F. Yuan and P. Huang, Class. Quantum Grav. 34, 077001 (2017).
 \bibitem{9}J. C. Fabris, M. H. Daouda and O. F. Piattella, Phys. Lett. B 711, 232 (2012).
\bibitem{10}H. Moradpour, Phys. Lett. B 757, 187 (2016).
\bibitem{sud0}S. Soroushfar, R. Saffari, S. Upadhyay, General Relativity and Gravitation 51 (2019) 130.
\bibitem{Ras1}J. C. Fabris, H. Velten, T. R. P. Carames, M. J. Lazo and G. S. F. Frederico, Int. J. Mod. Phys. D 27, 1841006 (2018).
\bibitem{Ras2} W. A. G. De Moraes, and A. F. Santos, General Relativity and Gravitation 51, 167 (2019).
\bibitem{kumar}R. Kumar and S. G. Ghosh. Eur. Phys. J. C 78, 750 (2018).

\bibitem{morad222}Iarley P. Lobo, H. Moradpour, J. P. Morais Graca, I. G. Salako, Int. J. Mod. Phys. D 27, 1850069 (2018).

\bibitem{das}
S. Das, P. Majumdar, R.K. Bhaduri, Class. Quantum Gravity 19, 2355 (2002).

\bibitem{rahim}
J. Sadeghi, B. Pourhassan, and F. Rahimi, Canadian Journal of Physics 92, 1638 (2014).

\bibitem{Hawking1}
S. W. Hawking, Nature 248, 30 (1974).

\bibitem{Hawking2}
S. W. Hawking, Comm. Math. Phys. 43, 199 (1975).

\bibitem{NPB1}
B. Pourhassan, M. Faizal, Nucl. Phys. B 913, 834 (2016).

\bibitem{sud1}
B. Pourhassan, S. Upadhyay and  H. Farahani, Int. J. Mod. Phys. A 34 (2019) 1950158.
\bibitem{sud2}
N.-ul Islam, Prince A. Ganai and  S. Upadhyay, Prog. Theor. Exp. Phys. (2019). In press.
\bibitem{sud3}
S. Upadhyay and  B. Pourhassan, Prog. Theor. Exp. Phys. 013B03 (2019).
\bibitem{sud4}
S. Upadhyay, Gen. Rel. Grav. 50, 128 (2018).
\bibitem{sud5}
S. Upadhyay, S. H. Hendi, S. Panahiyan and  B. E. Panah, Prog. Theor. Exp. Phys. 093E01 (2018).
\bibitem{sud6}
B. Pourhassan, S. Upadhyay, H. Saadat and  H. Farahani, Nucl. Phys. B928, 415 (2018).
\bibitem{sud7}
S. Upadhyay, Phys. Lett. B775, 130 (2017).
\bibitem{sud8}
B. Pourhassan, M. Faizal, S. Upadhyay and  L. A. Asfar, Eur. Phys. J. C 77, 555 (2017).
\bibitem{sud9}
S. Upadhyay, B. Pourhassan and  H. Farahani, Phys. Rev. D95, 106014 (2017).
\bibitem{hay}
B. Pourhassan, M. Faizal, and U. Debnath, Eur. Phys. J. C 76, 145 (2016).
\bibitem{HD}
Y. Heydarzade, F. Darabi, Phys. Lett. B 771, 365 (2017).
\bibitem{Kiselev}
V. V. Kiselev, Class. Quantum. Grav 20, 1187 (2003).

\bibitem{1908.09629}
H. L. Prihadi, M. F. A. R. Sakti, G. Hikmawan, and F. P. Zen [arXiv:1908.09629].

\bibitem{EPL}
B. Pourhassan, M. Faizal,  EPL 111, 40006 (2015).
\bibitem{PV}
J. Sadeghi, B. Pourhassan, M. Rostami, Phys. Rev. D 94, 064006 (2016).


\bibitem{G1}
G. Ruppeiner, Phys. Rev. A 20, 1608 (1979).
\bibitem{G2}
F. Weinhold, J. Chem. Phys 63, 2479 (1975).

\bibitem{G3}
P. Salamon, J. D. Nulton and E. Ihrig, J. Chem. Phys. 80, 436 (1984).
\bibitem{G4}
 R. Mrugala, Physica. A (Amsterdam), 125, 631 (1984).

\bibitem{higher1}
B. Pourhassan, K. Kokabi, S. Rangyan, Gen. Relativ. Gravit. 49, 144 (2017).
\bibitem{higher2}
B. Pourhassan, K. Kokabi, Z. Sabery, Annals of Physics 399, 181 (2018).
\bibitem{higher3}
B. Pourhassan, Eur. Phys. J. C79, 740 (2019).
\end{thebibliography}
\end{document}